\documentclass[14pt]{article}
\usepackage{jheppub}
\usepackage{amsmath,float}
\usepackage{epsfig,graphics,graphicx}
\usepackage{subcaption}
\input epsf
\def\be{\begin{equation}}
\def\ee{\end{equation}}
\def\barr{\begin{array}{lr}}
\def\earr{\end{array}}
\def\bea{\begin{eqnarray}}
\def\eea{\end{eqnarray}}
\def\nn{\nonumber}

\def\del{\partial}

\def\a{\alpha}
\def\b{\beta}

\def\d{\delta}
\def\e{\epsilon}

\def\g{\gamma}

\def\k{\kappa}
\def\l{\lambda}
\def\m{\mu}
\def\n{\nu}

\def\r{\rho}
\def\s{\sigma}

\def\D{\Delta}

\def\O{\Omega}

\title{\bf Thermodynamic Properties of Holographic superfluids}

\author{Roopa D'Almeida,} \author[b]{K. P. Yogendran\footnote{On
    deputation from IISER, Mohali}} \affiliation[b]{Indian Institute
  of Science Education and Research Tirupati\\ Karakambadi Road,
  Mangalam, Tirupati, Andhra Pradesh, India} \emailAdd{rooparda@gmail.com}
\emailAdd{pattag@gmail.com}

\abstract{Using the holographic model for spontaneous symmetry
  breaking, we study some properties of the dual superfluid such as
  the thermodynamic exponents, Joule-Thomson coefficient,
  Compressibility etc. Our focus is on how these properties vary with
  the scaling dimension and the charge of the operator that undergoes
  condensation.}

\arxivnumber{}
\begin{document}
\maketitle

\section{Introduction}

Following the paper of Gubser \cite{Gubser}, Hartnoll, Herzog and
Horowitz \cite{HHH} constructed a model exhibiting spontaneous
breaking of a U(1) gauge symmetry in the 5-d Einstein-Maxwell-Scalar
system with a positive cosmological constant {\em in the absence of
  a potential for the scalar field}. This phenomenon has an
interpretation, via the gauge gravity duality, as the spontaneous
breaking of a $U(1)$ global symmetry of a boundary theory that is dual
to the bulk system. The dual ``field theory'' corresponding to these
gravity systems contain condensates which are responsible for the
breaking of the global symmetry and an excitation spectrum that gives
rise to superfluidity (dissipation-less transport). Since then, much
fluid has flown under this holographic bridge. These include various
modifications of the initial action, the study of fluctuations,
hydrodynamics and extensions to non-relativistic cases, dynamics etc.
Reviews that summarize some of the developments are
\cite{Hartnoll:2009sz}-\cite{McGreevy:2016myw}. Much of the focus of
this subsequent work has been on the phases of such systems, transport
properties and some questions regarding entanglement entropy and the
dependence of these quantities on bulk parameters.

\subsection{Literature Review}
In this short survey, we attempt to collect together a list of
articles in which some bulk parameter is tuned under various
motivations. This list is not exhaustive and we welcome
suggestions and requests for citation.

Early on, \cite{Horowitz:2008bn} studied the variation of the
superfluid density, the gap and the order parameter with the scaling
dimension, but in the probe limit (in 3 and 4 boundary
dimensions). These authors also observed that the mass of the bulk
scalar $m^2$ plays the role of the interaction strength. In
\cite{Gubser:2008pf}-\cite{Horowitz:2009ij}, the zero temperature
ground state properties were studied. The work by \cite{SJS} showed
that the parameter space of solutions to gravity system is quite
involved showing the presence of numerous branches of solutions close
to each other. In \cite{Umeh:2009ea}, the dependence of the condensate
on the scaling dimension was studied albeit {\em in the probe
  approximation}, and a qualitative difference between $\D_-$ and
$\D_+$ condensates was argued for. The $\D_-$ condensate diverged at
low temperature, but that was an artifact of the probe limit
\cite{HHH2}. In the latter work, the dependence on $q$ was also
studied, where it was pointed out that $\D_-$ and $\D_+$ condensates
behaved differently as a function of $q$ - the bulk gauge coupling. It
was also pointed out that condensation occurs for even neutral scalars
and thus suggested a second mechanism driving the condensation due to
the formation of an $AdS_2$ throat ( distinct from the condensation of
charged scalars because the effective mass becomes tachyonic). In
\cite{Liu:2010ka,Pan:2011ns,Aprile:2009ai,Franco:2009if}, the bulk
gauge field was coupled to a Stuckelberg field and the phase diagram
and electrical conductivity were studied. \cite{Pan:2009xa} studied
the behaviour of the condensate in a system with Gauss-Bonnet
couplings in the bulk in the probe limit for varying Gauss-Bonnet
coupling and scalar field mass (and also varying spacetime
dimension). \cite{Brihaye:2010mr} improved the previous calculation
going beyond the probe limit and observed that the critical
temperature decreased with increasing GB-coupling (in 3+1-D). In the
work of \cite{Charmousis:2010zz}, a more detailed study of
thermodynamic properties such as heat capacity and the grand potential
were studied using a bulk system with a real scalar field (dilaton)
which triggers the symmetry breaking. In \cite{Arean:2010zw}, the
authors studied the onset of condensation for various values of the
mass of a bulk scalar interacting with a Maxwell field in the
background of a neutral black brane in d=3,4. The authors find an
interesting phase diagram in the superfluid velocity-temperature plane
as the dimension and mass of the scalar field is varied. In \cite{Way}
- the authors constructed a detailed phase diagram for various values
of $q$ by starting with a 3-D boundary system compactified on a circle
of radius $\g$ . In this case, there is another saddle point namely
the AdS soliton that competes with the black hole - and is interpreted
on the boundary as an insulating phase. An early suggestion in
\cite{KKY1,KKY2} that the bulk scalar mass can model the (Feshbach)
interaction strength in the dual system was reviewed in \cite{KKY}
where the dependence on the mass was detailed. In
\cite{Edalati:2010ge}, a bulk fermion coupled to the gauge field was
studied in the probe approximation. Varying the (dipole) coupling
strength produced interesting phenomenology vide the cuprates.
Another approach to BCS phenomenology using bulk fermions appears in
the work of \cite{Liu:2014mva}. In \cite{Dias:2011tj}, an exhaustive
study of the phase diagram of this system in $d=4$ with $m^2=0$ was
performed for spherical boundary topology, in particular varying the
charge of the scalar field. It was pointed out that the nature of the
ground state changed discontinuously as a function of the electric
charge. Bifundamental fields \cite{Arean:2015wea} have also been
considered with a view to obtain more interesting phenomenology The
work of \cite{Plantz:2015pem} focuses on the fluctuations of the order
parameter and studies variations with respect to the mass and
charge. Another modification of the bulk action, this time with two
scalar fields and quartic self-couplings is the work
\cite{Chen:2016cym} who also obtain a phase diagram as bulk parameters
are varied. Recently, \cite{deWolfe} introduce double trace
deformations of the charged scalar as modeling the interaction
between fermionic constituents of the boundary system towards
capturing the physics of the BEC-BCS crossover. This interaction leads
to a shift in the effective mass of the bulk scalar and thus is
suggestively similar to the earlier proposal in \cite{KKY1} that
tuning the mass of the bulk scalar is the way to go in capturing this
crossover.

As is seen in this survey, the emphasis has mostly been on
modifications of the bulk action and exploring the resulting phase
diagram and transport coefficients with a view to ``interesting''
boundary phenomenology. Our present study, on the other hand will
discuss thermostatic properties such as the energy, entropy, equations
of state and a few response coefficients of these systems by including
the backreaction of the charged matter fields on the gravitational
fields. We do not turn on additional potentials for the scalar field. 

The usual approaches to superfluidity involves using the
Gross-Pitaevski equations (order parameter equations) or the
microscopic Bogoliubov-de Gennes equations (in case of fermionic
constituents). A significant focus of study in these contexts has been
to explore the dependence on the scattering length $a_F$ of the
constituents - especially in view of the fact that $a_F$ can be varied
experimentally using magnetic fields (and atoms). The holographic
approaches are however significantly easier to handle and therefore a
question of interest is to ask if we can map microscopic parameters
(for instance $a_F$) onto the parameters in the gravitational
equations.

In this work we propose to treat the scaling dimension of the operator
that undergoes condensation as a proxy for the interaction strength
(this was initially explored using solitons in \cite{KKY1}-\cite{KKY2}
and reviewed in \cite{KKY}). In that work, it was observed that the
amount of condensation depends strongly on the scaling dimension. If,
for instance, the system consists of bosons - we may expect that the
condensate is of BEC type. On the other hand, if it is fermionic
matter that undergoes pairing followed by condensation, then at low
interaction strengths, the Cooper pairs are large and floppy, and we
may expect that the condensate exhibit properties of fermion bilinears
(and perhaps also show vestiges of a Fermi energy). We had presented
evidence in that work that solitons with $\D=2$ involve two length
scales similar to Fermi systems undergoing condensation. However, if
the interaction between the fermions is very strong, then the Cooper
pairs could be tightly bound and are effectively bosonic - thus the
condensate should be of the BEC type. We had shown that the ones with
$\D=1$ were more like Bose systems with quantitatively larger amounts
of condensate.

We thus expect to see significant differences in even thermostatic
properties such as entropy and the chemical potential on either side
of this crossover as we vary the scaling dimension of the operator
undergoing condensation and the electric charge of the scalar field
responsible for this phenomenon.

This document is organized as follows. In the first section we will
present the basic equations of the model in order to be
self-contained. We will then recall the symmetries of the action that
will allow us to scale the various quantities and simplify the
parameter space. This allows multiple interpretations of the same
classical solution; the various thermodynamic quantities obtained from
the {\em action} are however different.

The second section will briefly discuss some numerical issues that
arise in exploring the parameter space of the solution, now including
the backreaction of the scalar field.

The subsequent sections organize the new results as follows.


\section{Model}

The action we adopt in this work is
\be
S=\int
\frac{\sqrt{g}}{16\pi G_N}\left(R+\frac{2d(d-1)}{L^2}-\frac{1}{\l^2}
     [\frac{1}{4}F^2+ |D\psi|^2+V(\psi)]
\right)\label{action}
\ee
where we shall define the gauge covariant derivative
$D\psi=(\partial+i e A)\psi$. This action is to be supplemented with
several boundary terms (including the Gibbons-Hawking term) in order
to obtain consistent equations, 2-point functions and finite values
when evaluated on solutions. These boundary terms maybe obtained by
the procedure of holographic renormalization \cite{Skenderis}.

We will interpret $G_N\sim \frac{1}{N^2}$ with $N$ as the number of
colors, and $L$, having dimension length, sets the scale of the volume
in the field theory (and hence maybe scaled to unity).

Here, $e\lambda$ is the charge of the scalar field (as can be seen by
canonically normalizing the field $A$) and $\lambda$ controls the
interpretation of the gauge field $A$. For instance, $A$ can be
interpreted as baryon number if $\lambda \sim \sqrt{N}$.

Then an application of Gauss' law to a solution with a charged black
hole in the bulk of AdS will give the charge carried by the black hole
as being proportional to N (thus correctly taking care of deconfined
quarks - each baryon being regarded as being made up of N
quarks).

This interpretation is supported by regarding $F$ as the gauge field
on the world-volume of some compact branes which is a plausible
holographic model for baryons (in this case, the normalization of the
DBI action for the wrapped brane would account for the factor of $N$).
The other possibility is the interpretation of $A$ as a quark
number. In this case, we should set $\lambda\sim O(1)$. In this case,
the $F$ is interpreted as the world volume $B$-field of a fundamental
string ending on the horizon of the black hole. In both cases, we
assume a constant dilaton profile.

We can get rid of the charge $e$ by scaling the matter fields and
another scaling of the scalar field by a common factor $\lambda e$
results in an overall factor of $\frac{1}{q^2}=\frac{1}{e^2\l^2}$
multiplying the matter part of the action. We must however utilize the
action \ref{action} in order to determine field theory quantities. The
holographic duality suggests that we interpret the action, evaluated
on a given classical solution, as the generating function of connected
correlation functions of the dual theory (the sources are introduced
via boundary terms for the bulk fields). And since the latter are
obtained by taking functional derivatives with respect to the sources,
the coefficients of the various terms in the bulk action affect the
relative normalization of the various n-point functions.

We will assume that the bulk metric takes the following form 
$$ ds^2=- g\, e^{-\chi} dt^2+ \frac{dz^2}{g}+\frac{(dx^2+dy^2)}{z^2}$$
and that the solution involves only the scalar potential $\phi$. We
will look for solutions with a real profile for the scalar field
$\psi$ and all fields are assumed to be functions of the radial
co-ordinate z only.

The equations for arbitrary boundary dimension $d$ are 
\begin{eqnarray}
  \Psi''+\left(\frac{g'}{g}-\frac{\chi'}{2}-\frac{d-1}{z}\right)\Psi'
  +\frac{\phi^2 e^\chi}{g^2}\Psi-\frac{V'(\Psi)}{2g}=0 \\ \nn
  \phi''+\left(\frac{\chi'}{2}-\frac{d-1}{z}\right)\phi'-\frac{2\Psi^2}{g}\phi=0\\\nn
  \chi'-\frac{4}{z}-\frac{2z}{q^2(d-1)}
  \left((\Psi') ^2+\frac{\phi^2\Psi^2 e^\chi}{g^2}\right)=0 \\ \nn
  g'-\left(\frac{d}{z}+\frac{\chi'}{2}\right)g+dz
  -\frac{2z}{q^2(d-1)}\left(\frac{V(\Psi)}{2} +\frac{(\phi') ^2 e^\chi}{4}\right)=0
\label{eqn}
\end{eqnarray}

Recall that the system of equations above are invariant under two
independent transformations \cite{Gubser} - scaling of all the
coordinates by a common factor and rescaling only the time
coordinate. That is to say, if $g,h=e^\chi,\phi,\Psi$ solve the equations
then so will
\begin{equation}
\tilde g(z):=\a^2\, g(\frac{z}{\a})\hspace{1cm}
\tilde h(z):=\a^{2}\beta^2\, h(\frac{z}{\a})\hspace{1cm}
\tilde \phi(z):=\frac{1}{\beta}\, \phi(\frac{z}{\a})\hspace{1cm}
\tilde \Psi(z):= \Psi(\frac{z}{\a})\label{scaling}
\end{equation}

\section{Numerical solution and asymptotics}

To integrate the above equations numerically, we need to supply
various initial/boundary conditions. These are
\begin{equation}
\begin{aligned}
g(z=z_H)&=0 \hspace{1cm} \chi(z=0)&=& 0 \\
\phi(z=z_H)&=0 \hspace{1cm} \phi'(z=z_H)&=& \phi_0\\
\Psi(z=z_H)&=\Psi_0\hspace{1cm} \Psi'(z=z_H)&=& {\rm regular}
\end{aligned}
\end{equation}

The regularity condition on $\Psi'$ at $z=z_H$ is required in view of
the vanishing metric factor which indicates the presence of a black
hole in the bulk with the horizon at $z=z_H$. The freedom to rescale
time is eliminated by the condition that $\chi(z=0)=0$. Requiring
that the horizon appear at $z=z_H=1$ fixes the remaining coordinate scaling
symmetry of the equations.


Using the equations, we can see that the scalar field $\Psi$ has the
asymptotic expansion
\be
\Psi(z)\sim \Psi_n z^{\Delta_n} +... + \Psi_{nn} z^{\Delta_{nn}} \label{asymp}\ee
where
both exponents satisfy $\D(\D-d)=-m^2L^2$. The coefficient $\Psi_n$ of
$z^{\Delta_n}$ is interpreted as the condensate (i.e, VEV) of an
operator $\mathcal{O}$ in the boundary theory. This condensate breaks a global $U(1)$
symmetry manifests as a gauge symmetry in the bulk and is
therefore ``Higgsed'' by the scalar field profile. $\Psi_{nn}$ is then
an external source that turns on explicit symmetry breaking terms in
the boundary Hamiltonian. For {\em spontaneous} symmetry breaking, we
will tune the parameter $\phi_0$ to ensure that $\Psi_{nn}=0$.

In the range $\frac{d^2}{4}-1<-m^2<\frac{d^2}{4}$, either root of the
equation above maybe chosen as $\D_n$, whereas for larger values of
$m^2$, only the coefficient of the smaller power of $z$ can be
interpreted as the condensate \cite{KW}.

For instance, for $m^2=-2$, we may require a vanishing first (or
second) derivative at the boundary These two boundary behaviors
determine the scaling dimension of the operator that undergoes
condensation to be $\Delta_\pm=2,1$ respectively. Thus, the operators
maybe thought of as a fermion bilinear or a charged scalar bilinear in
the field theory language (recall that in d=3 a fermion has bare
scaling dimension 1 and a scalar field $\frac{1}{2}$).


We employ a simple Newton-Raphson iteration to determine that value of
$\phi_0$ for fixed $q$ and $\Psi_0$ which ensures that the solution of
the differential equation satisfies the boundary conditions on
$\Psi$. With a little experimentation, one can easily see that there
are multiple solutions for a fixed $\Psi_0$ (with different values for
$\phi_0$) differing by the number of nodes in the radial profile for
the scalar field.

In this work, we will restrict our attention to the solutions without
any node in the scalar field profile. The interpretation of the other
solutions with nonzero number of nodes is left for future work -
perhaps along the lines of \cite{Winters}. Some preliminary results
(obtained in joint work with Sudip Naskar (IIT Indore)) suggests that various
extensive quantities scale with the number of nodes. Following the
work of \cite{SJS}, who show that the space of solutions exhibits
strong sensitivity to the values of $q$, care must be taken to ensure
that one is exploring the correct branch of bulk solution. In
particular, for small values of $q$, there are multiple solutions for
the scalar field profile with no nodes. In this work, we have not
explored these other regions of parameter space.

The solution so obtained still has two independent parameters namely
$\Psi_0$ and $z_H$, which allows us to vary the temperature and
chemical potential independently. However, the scaling symmetry
\ref{scaling} of the equations of motion implies that we can keep one
of these fixed while obtaining numerical solutions without loss of
generality.

If one imagines that the bulk description is a kind of Ginzburg-Landau
free energy, then the parameters $z_H$ and $\Psi_0$ maybe expected to
have definite scaling properties with the temperature and chemical
potential at least close to $T_c, \mu_c$. This is indeed observed to
be the case.

The numerical solution is obtained by integrating out from the horizon
which maybe set to $z_H=1$. The solution so obtained has
$\chi(0)=\chi_0\neq 0$. We then employ the freedom to rescale the time
coordinate at the boundary to shift this boundary value of $\chi$ to
zero which rescales the chemical potential and the temperature. We may
then use the $\alpha$ rescaling \ref{scaling} to set either the
temperature or the chemical potential to any desired value.

A consequence of the scaling symmetry is that we have an equation of
state $E=(d-1)PV$ - this is because the trace of the stress tensor
vanishes due to this symmetry.

The adapted AdS/CFT correspondence suggests that the
thermodynamic quantities of the field theory are read off from the
boundary data of the bulk fields. Specifically, the chemical potential
$\mu$ and the number density $\rho$ of the boundary theory are read
off from the asymptotic behaviour of the bulk gauge field $A_0$ as
$$ A_0(z)\sim a_0+ ... + Q z^{d-2}+...$$ as $\mu=\frac{a_0}{q}$ and
$\rho={Q}{q}$. It maybe noted that the latter quantity is related to
the electric field at the boundary; hence, the particle number is the
electric flux at the boundary. Requiring that the potential vanish
\cite{Gubser} at $z=z_H$ thus determines an equation of state of the
boundary theory - i.e, the dependence of the number density
$\rho(\mu)$.

The boundary stress tensor is computed by determining the
coefficient of $r^d$ in the Fefferman-Graham (FG) form for the bulk
metric (this is sufficient in d=3)\cite{Skenderis}.

We provide a table summarizing the manner in which the thermodynamical
quantities of the system living on the boundary are read off from the
bulk fields.
\begin{center}
\begin{tabular}{|l|l|l|}\hline
  Entropy density & s&$\frac{1}{4G_Nz_H ^{(d-1)}}$\\\hline
  Temperature & T & $\frac{1}{4\pi}\frac{g'(z_H)}{\sqrt{h(z_H)}}$ \\\hline
  Chemical potential &$q\mu$&$a_0=A_0(z=0)$\\\hline
  Number density & $q\rho$ & $A_0 '(0)/16\pi G_N$ \\\hline
  Energy density & $\e$& $-\frac{(d-1) g_3}{16\pi G_N}$\\\hline
  Pressure & P& $-\frac{g_3}{16\pi G_N}$\\\hline
  Condensate & $\langle {\mathcal O}\rangle$ & $k(2\D-d)\psi_n$\\\hline
  \end{tabular}
\end{center}
From the equations it is clear that if $A_0$ is a solution then so is
$-A_0$. Choosing solutions so that $\mu<0$, it turns out that $\r>0$
(presumably to ensure that the boundary condition $A_0(z_H)=0$ on the
bulk gauge field is satisfied even though the latter quantity is
determined at the boundary and not at the horizon). This is consistent
with the idea that the chemical potential is negative for Bose
Einstein Condensates.

We can also observe that the electric flux increases monotonically out
to the boundary implying that the electric charge of the scalar cloud
is of the same sign as that of the black hole. Thus, electric
repulsion between the cloud and the black hole can counteract the
gravitational attraction leading to an equilibrium.

We also note that the actual chemical potential and number density can
be chosen to depend on $q$ in several ways depending on the
interpretation of the boundary current dual to the bulk gauge field
$A$ and the scalar field $\psi$. We shall return to this normalization
issue when we consider the variation of the thermodynamics with $q$.

The FG form for the bulk metric is $ ds^2=\frac{dr^2}{r^2} +
\frac{g_{ij} dx^i dx^j}{r^2}$. Assuming that the function $g$ in the
metric has a series expansion $g(z)=z^2(1+g_2z^2+g_3z^3+...)$ - a
similar series needs to be assumed for $h=e^\chi$ - we can perform a
coordinate transformation from the $r$ to the $z$ coordinate to
convert the metric to the FG form.

Inverting the resulting series to find $z(r)$ gives
$$ \frac{1}{z^2}=\frac{1}{r^2}\left(1-\frac{g_2 r^2}{2}-\frac{g_3
  r^3}{3}+(\frac{g_2 ^2}{16}-\frac{g_4}{4}) r^4+...\right).$$ Thus, the
boundary stress tensor along the x-directions is given by
$T_{xx}=-\frac{d}{4\pi G_N} \frac{g_3}{3}$ and the energy density
$\e=T_{00}=-\frac{d}{4\pi G_N}\frac{2g_3}{3}$ and hence boundary stress tensor
is traceless implying the equation of state $\e=(d-1)p$. 

By using the differential equations \ref{eqn}, and substituting
in the asymptotic expansion, it is easily determined that
$g_2=\frac{\Psi_1 ^2}{2 q^2}$ and $h_2=\frac{h_0\Psi_1 ^2}{2 q^2}$
where $\Psi_1$ is the value of the condensate in the case that $\D=1$
operator condenses.

In the case the operator that condenses has $\D=2$, the metric
expansion is $g_4=\frac{\Psi_2 ^2}{q^2}$ and $h_4=\frac{h_0\Psi_2
  ^2}{q^2}$ is the first non-vanishing coefficient.

We attempted to determine the coefficients $g_n$ by fitting the
numerical solution to a polynomial. The fit was required to be stable
against perturbation in the range over which the polynomial is
required to match the solution and also by perturbing the degree of
the polynomial. The goodness of fit was determined by the
(unnormalized) variance of the fitting function with respect to the
numerical solution. In particular, we first check that the
coefficients that ought to vanish are indeed numerically ``very
small'', that $g_{2,4}, h_{2,4}$ are indeed determined by the
condensate (as above from the expansion) and further that if the
fitting polynomial is modified by dropping the vanishing powers (using
the expansion as above), then the goodness of fit improves.

However, we find that the energy values are quite unstable in the
sense that varying the degree of the fitting polynomial and the cutoff
produce large changes in the numerical coefficient $g_3$. Therefore,
we adopted a different approach based on the Euler relation. Recall
that the extensivity properties of the energy of a thermodynamic
system implies that the state variables obey the Euler relation
$$ \epsilon=Ts-\mu \rho -p$$ with $s$ being the entropy density,
$\epsilon$ being the energy density and $\rho$ being the particle
number density.

The entropy of the boundary theory is defined to be the
Bekenstein-Hawking entropy of the black hole, namely
$S=\frac{A}{4G_N}$ where $A$ is the area of the horizon. In contrast
to the other observables of the boundary theory, entropy is determined
by a quantity evaluated deep in the bulk geometry. The temperature
was determined as the deficit angle in the Euclidean metric as
$T=\frac{1}{4\pi}\frac{g'(z_H)}{\sqrt{h(z_H)}}$.

\begin{figure}[H]
  \centering
  \includegraphics[height=6cm]{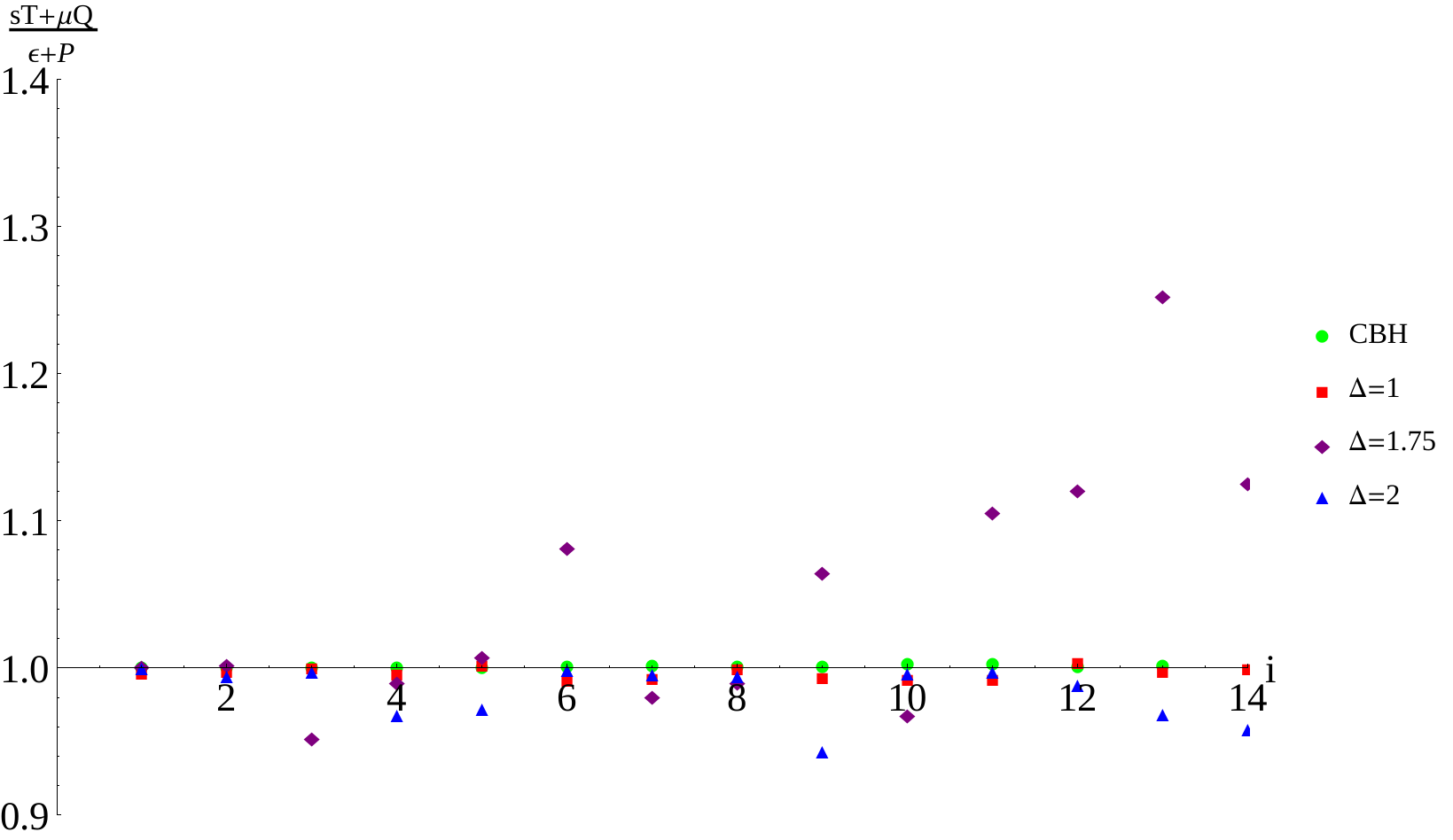}
  \caption{Verifying Euler relation: the horizontal axis is a solution index}
  \label{Euler}
\end{figure}

Using these values for the entropy and temperature, and the other
quantities we first verify that the Euler relation indeed holds for a
few scaling dimensions and over some range of temperatures by ensuring
that a stable and accurate energy value was determined (using the
above procedure). A graph of the ratio Fig:\ref{Euler} shows that
whenever the numerics are reliable, this thermodynamic consistency
condition is indeed obeyed.
Hence, in what follows - we simply determine the energy numerically by
using the Euler relation and the conformal equation of state
$E=(d-1)PV$ as $\epsilon=\frac{d-1}{d}(Ts-\mu \rho)$.

The action of the bulk theory when evaluated on the solutions is
interpreted as a thermodynamic potential of the boundary theory.
However one must add counterterms (living entirely on the boundary
thereby not affecting the equations of motion) to ensure that the
divergences coming from integrating upto $z=0$ cancel. A simple
approach is to use a minimal subtraction scheme which ensures
cancellation of only the divergent terms. A more sophisticated
approach is to use the method of ``holographic renormalization''
\cite{Skenderis}.

We find that neither of these approaches is numerically tractable for
similar reasons as mentioned in the evaluation of the energy. The
integration of the gravity part of the action turns out to be
extremely sensitive to the cutoff and ensuring that the numerical
values are insensitive to the choice of cutoff proved well-nigh
impossible.

However, not all is lost - using the Euler relation above the grand
potential is simply the negative of the pressure which can be
determined directly from the energy by using the equation of state
relating the pressure to the energy density. 

Since the particle number is computed by using Gauss' law, we can
calculate the number of particles carried by the black hole alone by
evaluating the electric flux at the horizon. In the dual theory, this
maybe interpreted as the number of particles occupying the excited
states of the single particle spectrum (although this may not be
sensible when the particles are strongly interacting). From the field
theory point of view, the interpretation of the condensate as the
occupancy of the ground state of the single particle spectrum has been
a source of some tension because of thermal fluctuations (and
especially interaction effects). This ambiguity in the separation of
the total particle number is also visible in the holographic
dual. Assuming the field theory dual to live at the boundary, it is
unclear if the electric field at the horizon (located in the interior
of the asymptotic spacetime) is an {\em independent} observable of the
boundary theory. Due to quantum effects, particles will be exchanged
between the scalar field and the black hole through ``Hawking
radiation'' and infall similar to what happens in the two-component
model of a superfluid. The electric flux at the boundary of AdS
however suffers from no such ambiguities and is directly interpretable
as the total number of particles.


\section{Entropy and the microcanonical ensemble}

We start by discussing the behaviour of the entropy of the system as a
function of the extensive variables: the total energy $E$ and the
total number of particles $N$ - namely, the fundamental thermodynamic
relation. The energy and the number of particles are the natural
variables characterizing the system in the microcanonical
ensemble. Note that the field theory coordinates $x,y$ span an
infinite volume in units of the radius $L$ of AdS (which has been set
to unity). Hence, the appropriate variables are the energy density
$\e=E/V$ and the number density $\r=N/V$.

Firstly, we shall identify the space of solutions of gravity system
corresponding to various thermal states of the dual field theory. In
the absence of the condensate, we have a thermal gas of particles in
the field theory which is described by a charged AdS black hole
\cite{Chamblin}, with the energy density $\e$

\be \epsilon=
\frac{(d-1) }{4\pi}(4 G_N s^d)^{\frac{1}{d-1}}(1+\frac{(d-2)}{32q^2(d-1)G_N ^2}
\frac{\rho^2}{s^2}) \label{F1}
\ee
exhibiting two different power laws at either extremes, depending on
the ratio dimensionless $\frac{\rho}{s}$. What is not obvious from
this expression is that the density is bounded $0<\rho<\rho_M$. From
the gravitational point of view, this is due to the extremality limit
- it is not possible to 'overcharge'  ($M_{bh}<Q_{bh}$ in appropriate units)
a black hole. 

\begin{figure}[H]
  \centering \includegraphics[height=6cm]{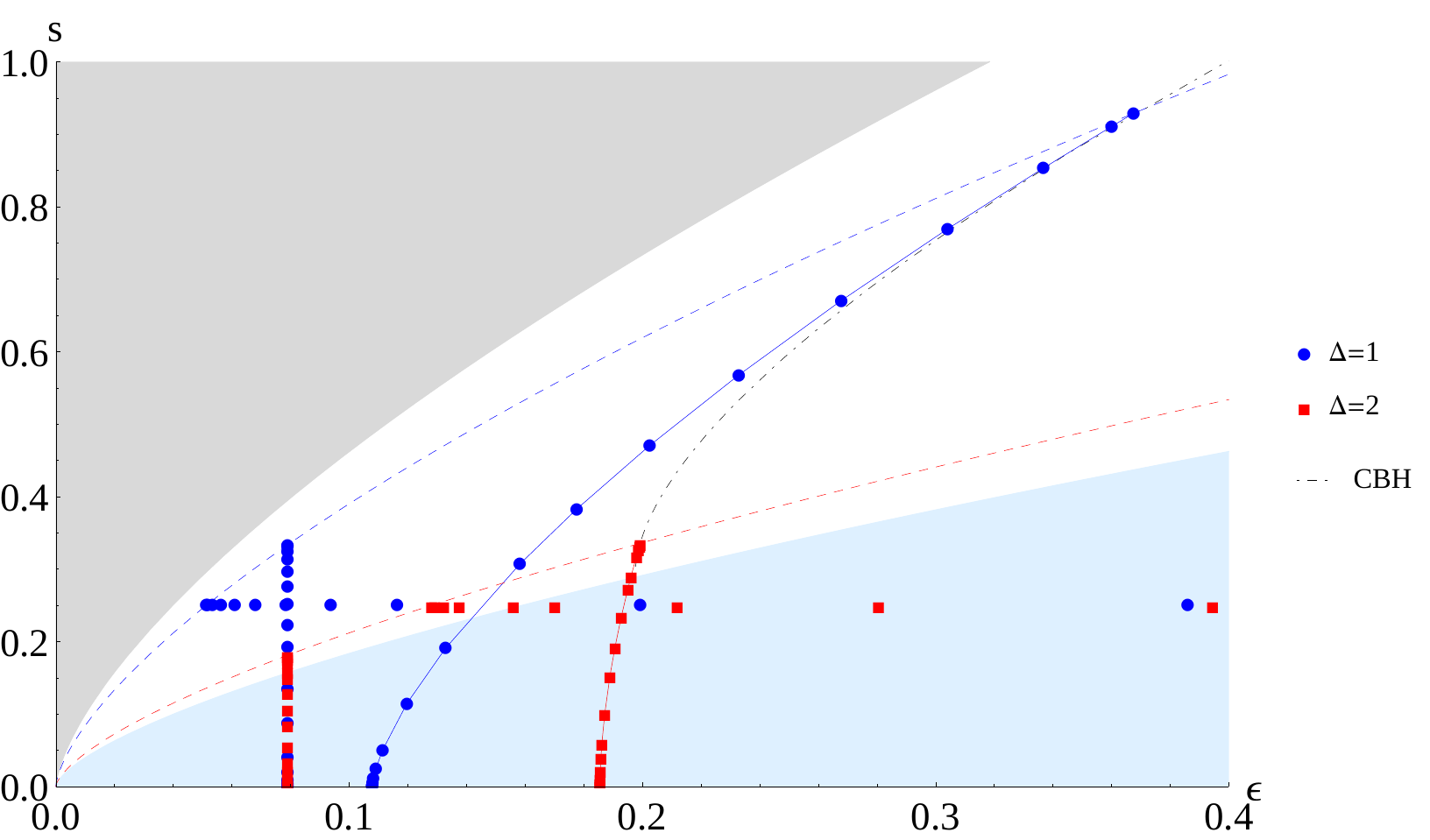}
  \caption{Entropy density vs Energy density}
  \label{Fig-SE1}
\end{figure}

These solutions fill out the region in white in Fig:  \ref{Fig-SE1}
where the dot-dashed line in black shows a family of charged black hole
(CBH) solutions of varying energy at fixed density. In the figure,
the upper boundary of the white region corresponds to $\rho=0$,
corresponding to neutral black holes. The lower boundary of the white
region corresponds to extremal black holes $\rho=\rho_M$ without a
condensate, that is to say charged black holes with $T=0$. Below this
region, there are no black hole solutions without a scalar field
component. From the point of view of the dual system, the extremal
black holes are a bit of a puzzle since they have nonzero entropy at
zero temperature. This violation of Nernst' law implies that these
black holes cannot be the correct description of the phase of this
system at low entropies.

On the other hand, for sufficiently large density (or chemical
potential), we have solutions with a condensate - represented on the
gravity side by solution with a nonzero profile for the scalar field.
These solutions fill out the region in blue in the figure. Two such
families of solutions are shown as a pair of solid curves in red and
blue at the same total number density as the black dot-dashed
curve. The line in blue represent solutions with a $\Delta=1$
condensate and a black hole while the line in red involves the
$\Delta=2$ condensate (all for the same total particle density). It is
seen that the condensate solutions "continue" the black hole curve
upto $s=0$. It is evident that the onset of condensation occurs much
earlier in energy for $\Delta=1$ than for $\Delta=2$. The figure, Fig:
\ref{Fig-SE1} suggests that entropy and the slope (i.e., temperature)
are both continuous, which would imply that the condensation
transition is continuous.

In the figure the row and column of (blue and red) dots, represent
points with the same value of $\rho$ but varying condensate value. The
blue and red dots represent condensation of operators of scaling
dimension $\Delta=1$ and $\Delta=2$ respectively. The uppermost (and
leftmost) points correspond to the onset of condensation. Note that
the onset of condensation does not extend into the upper grey region
of neutral black holes as expected. For the charge cloud to be stable
and located away from the black hole, the gravitational attraction of
the black hole needs to be cancelled by a repulsive force. If the
black holes are neutral, this is not possible, and the cloud will fall
in (thus, one can have different kinds of repulsive forces such as
dilaton charges leading to other interpretations on the boundary).

Starting from any one of these solutions, we may use the scaling
property \ref{scaling} to obtain new solutions. These scaled
solutions fall on the dashed curves in red and blue in the figure
(these were obtained by scaling the leftmost red and blue point with
$s=1/4$). It maybe noted that the curves do not cut across regions of
different color.

The following conclusions may then be drawn from a study of the space
of solutions in the $s-E$ plane (the points listed below apply to the
situation where the scaling dimension of the operator undergoing
condensation is fixed
\footnote{In the context of string theories and
  supergravity, there are typically families of (complex) bulk scalar
  fields with varying masses including multitrace versions of single
  trace operators. Thus, it is quite conceivable that more than one
  type of condensate is present in a given phase \cite{deWolfe}}
):

\begin{itemize}
\item
 For every point in the blue region, there is a black hole condensate
 solution without a node in the scalar field profile.
\item
For every point in the white region, there is either a unique charged
black hole solution with no condensate, or

There are two solutions with black holes, one of them with a condensate. 

\item
  For any $q$ and $\Delta$, and for constant particle number $\r$, at
  high entropy we have only charged black hole solutions without any
  condensate. At low enough entropy, solutions with condensates start
  to appear. There is a small window of entropy and energy where both
  solutions co-exist after which the charged black hole curve
  terminates at an extremal charged black hole. However, solutions
  with a condensate continue further upto $s=0$ (see the solid curves
  in Fig: \ref{Fig-SE1}).

 

 


\item
It is also seen that for a fixed entropy, $s=1$ say, the solutions
with a condensate exist beyond a critical value of energy
$\e_c$. These solutions are represented by horizontal sequence of blue
and red dots in figure Fig: \ref{Fig-SE1}. It maybe noted that the
$\Delta=1$ dots (in blue) start closer to the neutral black hole
boundary (grey region) compared to the $\D=2$ dots (in red) that start
closer to the extremal black hole i.e., critical value $\e_c$ at which
condensation starts increases with $\Delta$.

\item
Conversely, for a given value of $E$, there exist condensate solutions
below a critical value $s_c$ of entropy. The critical value decreases
with the scaling dimension and the temperature is correspondingly
lowered. These are demonstrated by the vertical sequence of points
marked in blue and red in the Fig: \ref{Fig-SE1}.

\item
Since the curves ``end vertically'' the slope $\frac{\del s}{\del
  \e}=\infty=\frac{1}{T}$ verifying Nernst' law. For a given value of
the number of particles $\rho$, the larger values of $\Delta$ have
larger energies at $T=0$ which is the intersection point of the
constant $\rho$ curves with the E-axis.

\item
A plot of the entropy vs {\em temperature} at constant particle number
is shown in Fig: \ref{sTQ} where the green curve represents the
charged black hole solutions. The solutions with a condensate branch
off at different values of $T_c$ for the different scaling dimensions
- the lower values condensing earlier. 

This figure shows that upon the inclusion of the condensate solutions,
the entropy of the system smoothly goes to zero at absolute zero
temperature satisfying the Third law of Thermodynamics (which the
charged black holes by themselves fail to obey). The location of the
horizon of the black hole moves deeper into the bulk of AdS at lower
temperatures thereby decreasing the entropy, while the $T=0$ system
has no black hole.
\item
Since entropy and temperature are continuous at the onset of
condensation - this transition is smooth without any latent
heat. However, the derivative of the entropy jumps at condensation -
implying a discontinuity in the specific heat.

Because of the underlying conformal symmetry, we have an equation of
state $p=\e/2$ ($d=3$) which implies that $\e_0$ is also the
degeneracy pressure (i.e., pressure at zero temperature).

\begin{figure}[h]
\flushleft
\begin{minipage}{.5\textwidth}
  \centering \includegraphics[width=\linewidth]{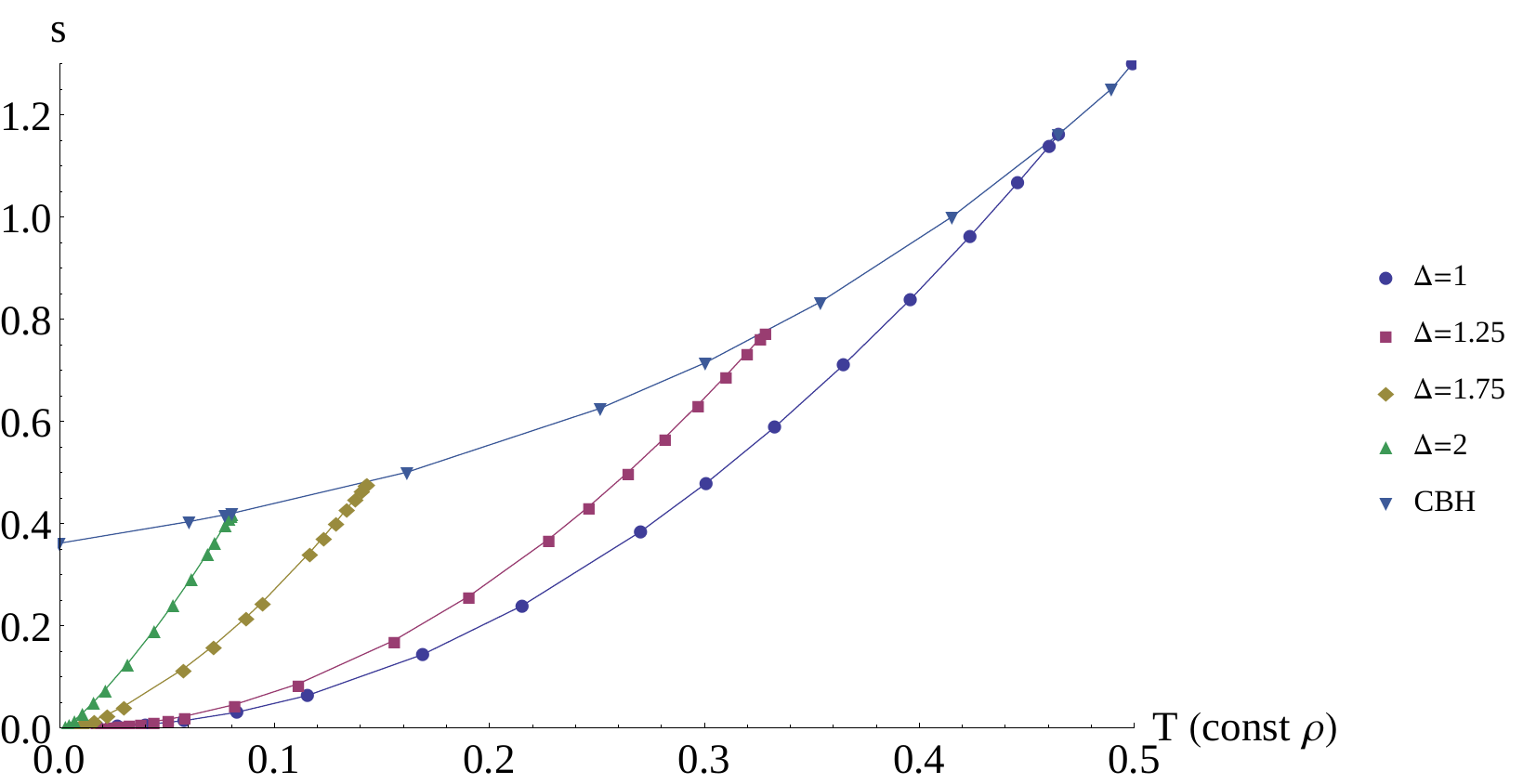}
  \captionof{figure}{Entropy vs. T at fixed $\rho$}
  \label{sTQ}
  \end{minipage}%
\begin{minipage}{.5\textwidth}
  \flushright
    \includegraphics[width=\linewidth]{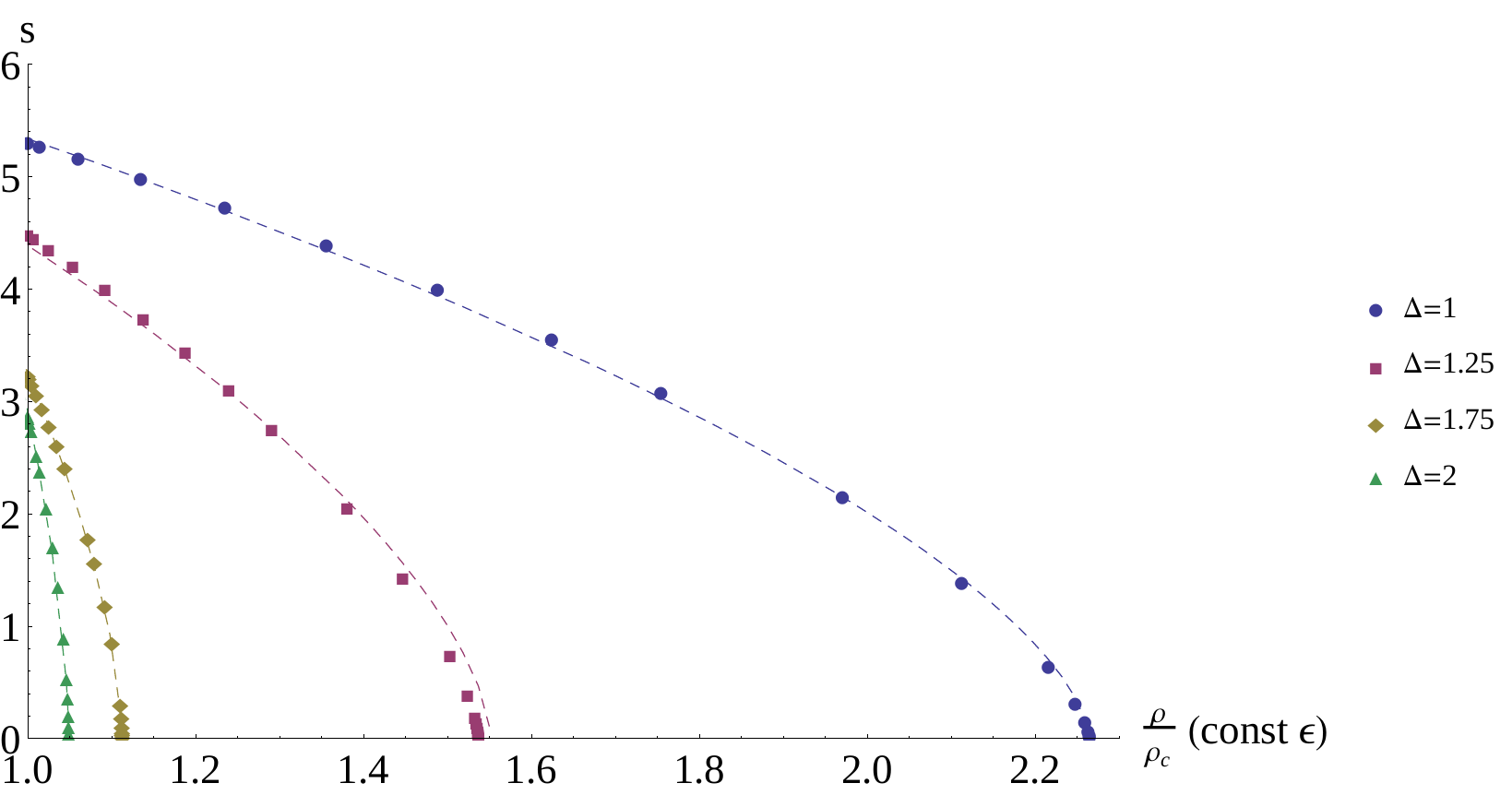}
  \captionof{figure}{Entropy vs. $\r$ at fixed $\e$}
  \label{sQe}
\end{minipage}
  \end{figure}
\item

Plotting the entropy (Fig: \ref{sQe}) as function of the density
holding the energy fixed again shows that the entropy goes to zero
smoothly as $T$ goes to zero (the graphs ``end vertically'' on the
x-axis).

\item
 At a given energy (for the same number density), once condensation
 occurs - the solution with the condensate always has higher entropy
 numerically and is thus the preferred phase (for all values of $q$
 and $\D$). This is seen by comparing the dot-dashed curve with the
 red (or blue) solid curves in Fig: \ref{Fig-SE1} (the latter
 correspond to condensates with $\D=2,1$ respectively)- for any given
 energy, the solid curves are always above the dot-dashed curve.

In gravitational terms, this statement translates into saying that the
area of the (charged) black hole increases once the cloud of charge
represented by the scalar field appears (at fixed number density for
low enough energy or for high enough density at fixed
energy). Lowering the total energy of the system is achieved by moving
charge out of the black hole into the scalar cloud surrounding it -
assuming that the cloud models ground state occupancy in the usual
picture of Bose condensation. This process seems to have a smaller
electrical effect than gravitational (then, the horizon moves closer
to the boundary, decreasing $z_H$ thereby increasing the
entropy). This conclusion is not obvious since one could have expected
that the small charge on the cloud can only repel the black hole
pushing the horizon further into the bulk.

\item
The power-law behaviour of energy as a function of temperature is
shown in Fig: \ref{ecv} - the lower curves represent the specific
heat (scaled by a factor of ten) which is the derivative of the energy.
  \begin{figure}[h]
 \centering
\includegraphics[height=6cm]{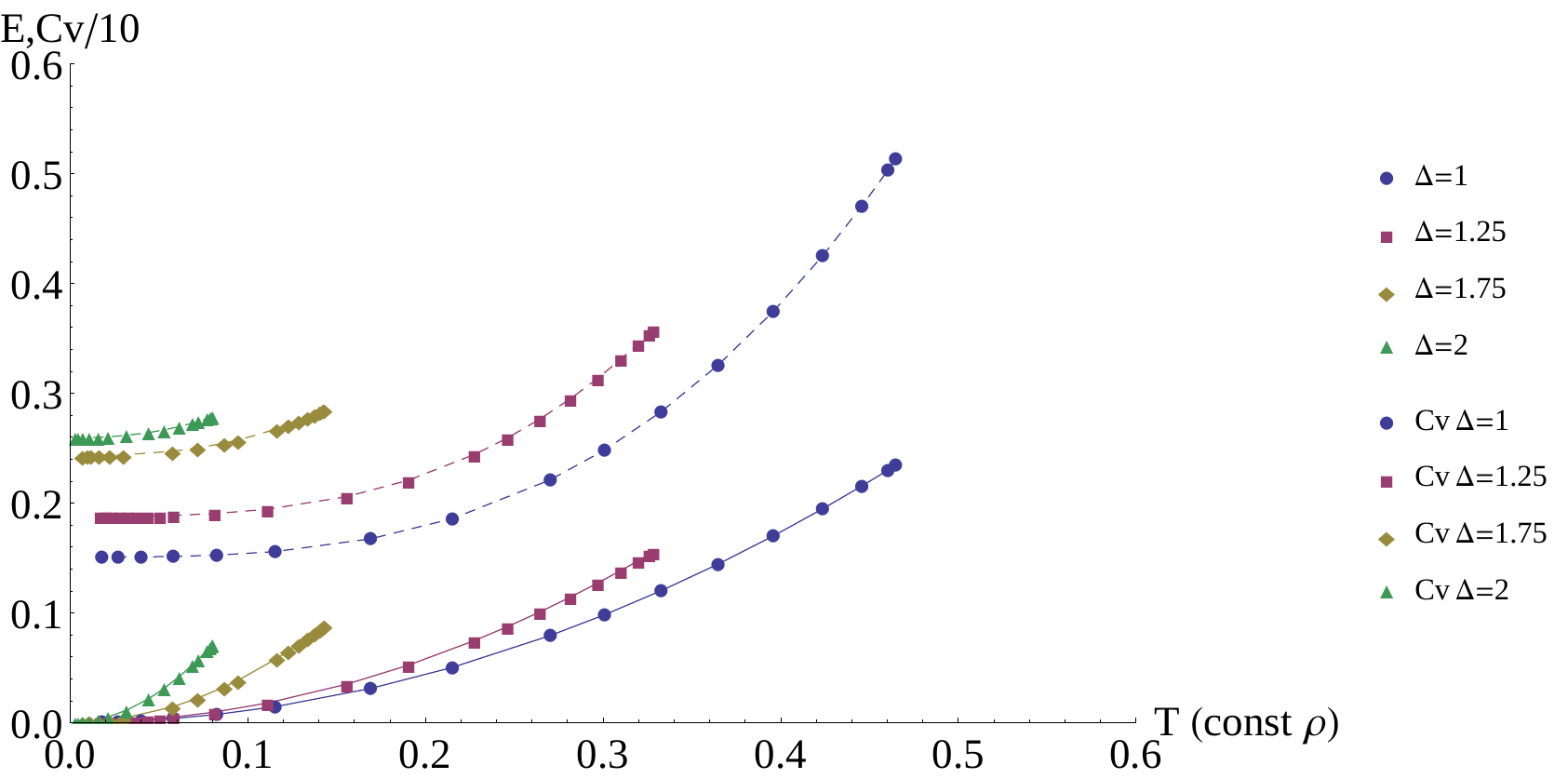}
\caption{\label{ecv} $\e, \frac{C_V}{10}$ vs T for constant $\r$}
\end{figure}
The discontinuity in the specific heat maybe seen in the graph of the
energy density (at constant number density) Fig: \ref{ecv} - the
finite discontinuity in the specific heat is markedly different from
the lambda transition of superfluid liquid helium.
\item
As shown in Fig: \ref{muqe}, the chemical potential turns out to be
linear function of the charge density. In the absence of the
condensate, i.e., for the charged black hole alone, the bulk gauge
field is a linear function of the holographic direction $z$, and
imposing the boundary condition that $A_0(z_H)=0$ gives a linear
relation between $\mu$ and $\r$ (see Appendix). Once the scalar field
is present, the gauge-field profile in the z-direction is no longer
linear. The nearly linear behaviour of the chemical potential is
surprising because in the bulk language this means that the scalar
cloud hardly affects the bulk gauge field profile.  Of course, the
equivalent statement in the usual understanding of condensation is
that the chemical potential contribution of the condensate is zero.
\begin{figure}[H]
\flushleft
\begin{minipage}{.5\textwidth}
  \centering
  \includegraphics[width=\linewidth]{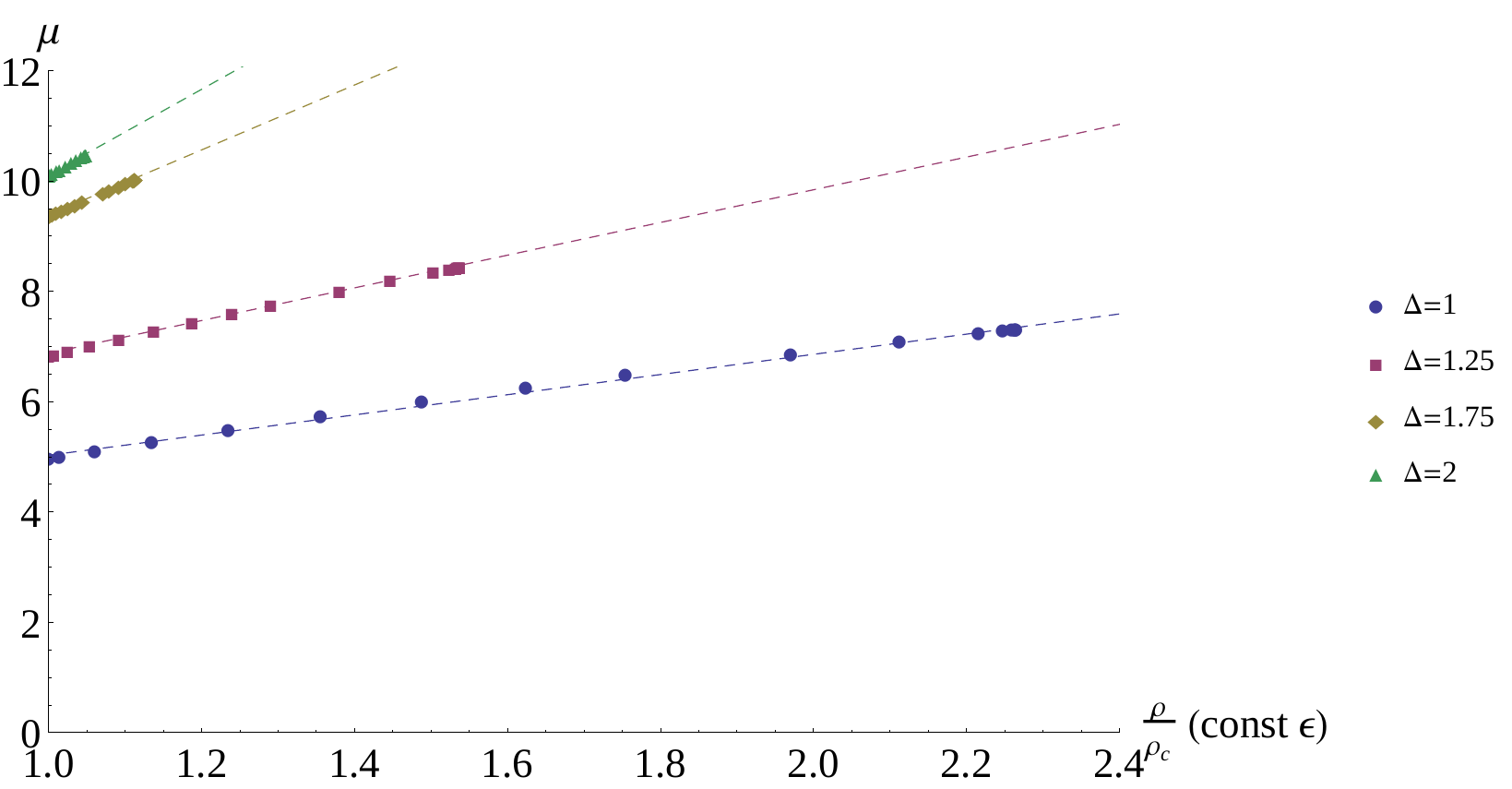}
  \captionof{figure}{$\mu$ vs $\rho$ for constant $\e$}
  \label{muqe}
  \end{minipage}%
\begin{minipage}{.5\textwidth}
  \flushright
   \includegraphics[width=\linewidth]{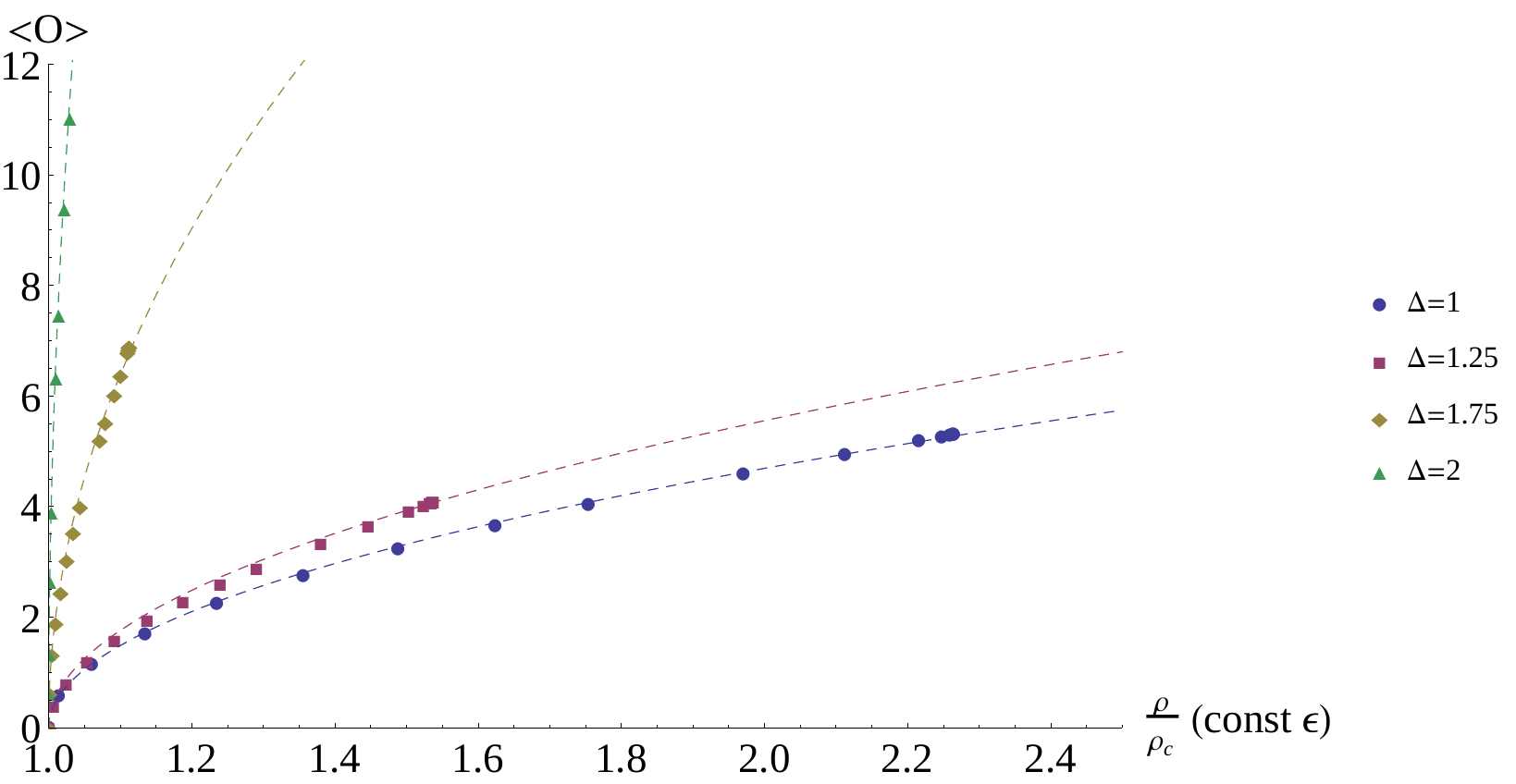}
  \captionof{figure}{VEV vs. $\r$ at fixed $\e$}
  \label{oqe}
\end{minipage}
  \end{figure}
Considering the range that the data cover in Fig: \ref{muqe} from the
onset of condensation at $\rho=\rho_c$ to $T=0$, the density varies by
a factor of two for small $\D$, whereas for the larger $\D$, the
density hardly changes.

A similar statement obtains for the energy variation suggesting that
for larger scaling dimensions, most of the states undergoing
condensation at finite temperature are from occupied states near the
ground state.

\item
To conclude, we show the behaviour of the condensate (order parameter
VEV) in Fig: \ref{oqe}; the numerical data can be fit by a functional
form $\mathcal{O}=\mathcal{O}_0\sqrt{\frac{\r}{\r_c}-1}$. We shall
return to this in section \ref{delta}. However, the large difference
as we vary the scaling dimension is noteworthy.
\end{itemize}

\subsection{Entropy and condensation}

\begin{figure}[H]
\flushleft
\begin{minipage}{.5\textwidth}
\includegraphics[width=0.8\textwidth]{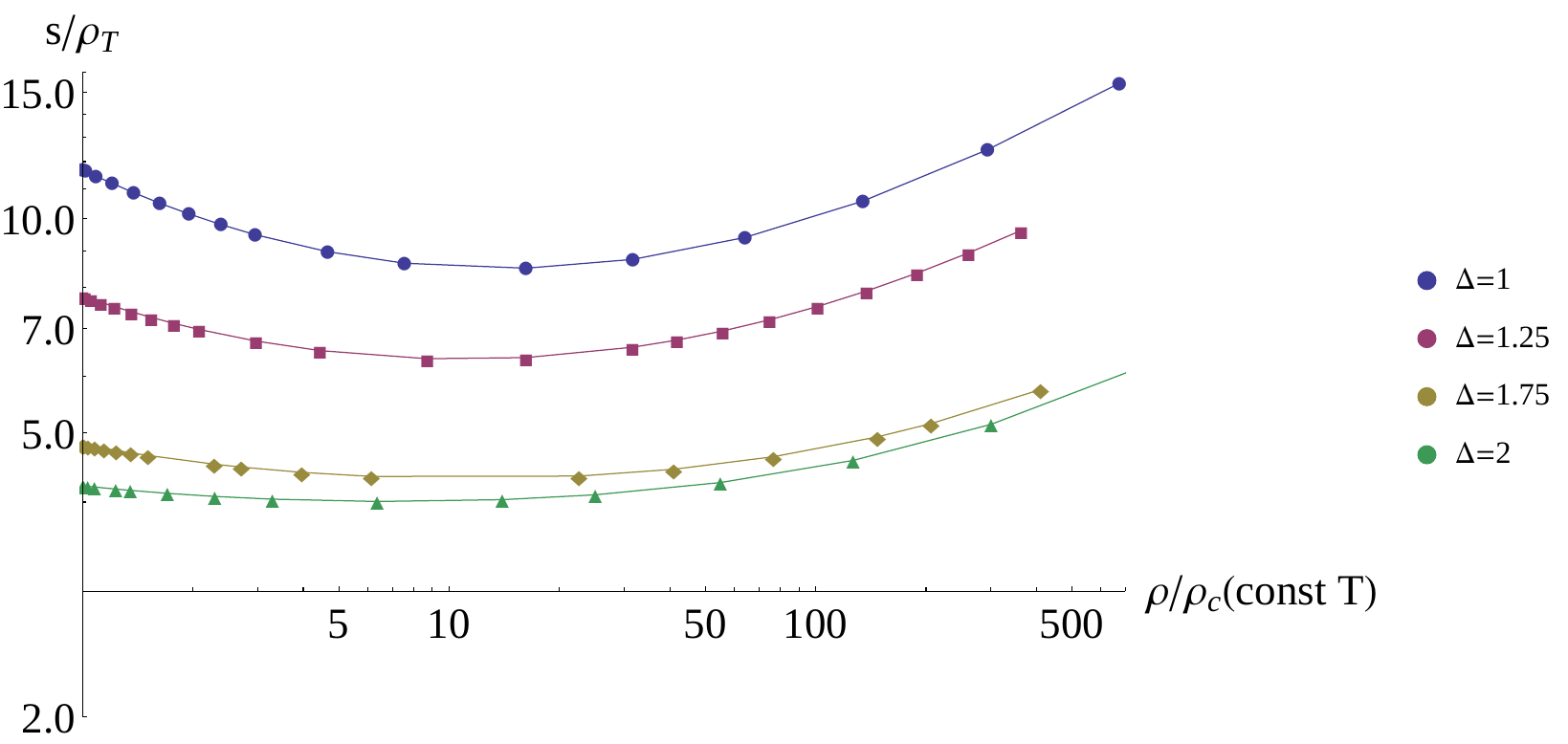}
  \caption{s vs. $\rho$ at fixed low $T$}
  \label{sqT}
  \end{minipage}%
\begin{minipage}{.5\textwidth}
  \flushright
  \includegraphics[width=0.8\textwidth]{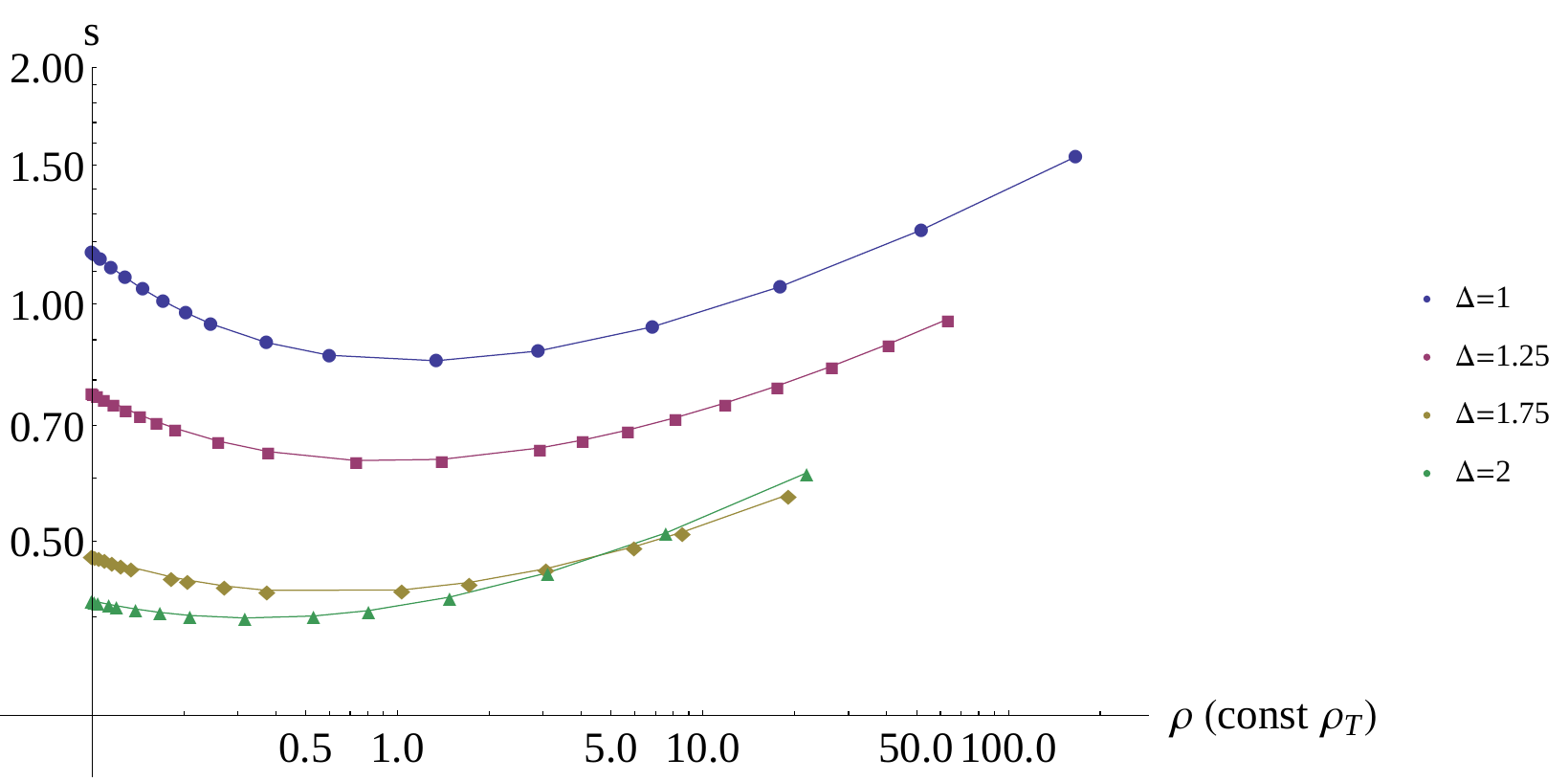}
  \caption{s vs. $\rho$ at fixed $\rho_{T}$}
  \label{sbyrho}
\end{minipage}
  \end{figure}

Once we know that condensation occurs, we can plot the entropy of the
system as a function of the total number density (at fixed
temperature). We may expect that, if the system were bosonic, the
entropy saturates because the added particles occupy the ground
state. This is shown in Fig: \ref{sqT} - where we normalize the
entropy by the charge on the black hole $\r_T$. This figure shows that
the entropy varies but little - most of increase in the number density
is added to the condensate which indeed contributes little to the
entropy.

The charge in the scalar cloud resides close to the horizon for larger
scaling dimension system- whereas for the smaller, the charge is more
distributed (see section \ref{soln-profile}). This is easily
understood from a Frobenius analysis of the bulk equations.  In the
holographic dictionary, the radial direction corresponds to energy
scales of the dual system. This means that the particles in the
condensate are, on the average, more dispersed in energy for smaller
scaling dimension as might be expected.

In the usual AdS holographic dictionary, the entropy of the boundary
theory is mapped to the area of the horizon of a black hole that is
present far in the interior of AdS space. However, this horizon area
when ``pulled back'' onto the boundary (so that it is a boundary
observable) could be expected to be ``distorted'' or lensed by the
energy density present in the scalar field profile in the intervening
region. If this happens, this is the gravitational interpretation of
entropy production/reduction. Thus, it is interesting to ask whether
the entropy is affected by the presence of the condensate. The field
theory interpretation would suggest that the entropy remains nearly
constant (the interaction (thermal loops) between the condensate and
the thermal cloud could alter this expectation). In the AdS side, we
therefore vary the total electric charge while keeping the charge on
the black hole fixed (thus only the number density in the condensate
changes).

In Fig: \ref{sbyrho}, we see that the entropy density remains nearly
constant over a large range of total charge (for fixed charge on the
black hole which we take to represent the thermally excited population
$\rho_T$).

The minima in this fixed {\em temperature} graph shows that for low
electric charge in the condensate cloud, the Coulomb repulsion leads
to decrease in horizon area (the horizon is 'repelled' from the
boundary), (Fig: \ref{sbyrho}) - but clearly gravitational attraction
dominates at large condensate density since the entropy eventually
increases. Note that, for larger values of the scaling dimension, the
minima seems to be disappear.

It will be interesting to understand the minima entirely from the dual
field theory perspective. Gravitational attraction and Coulomb
repulsion in terms of field theory?

This has the interpretation of a thermodynamic process which changes
the number of particles in the condensate while keeping the number of
particles in the thermal cloud fixed. This process deserves to be
studied better perhaps along the lines of \cite{Sonner:2014tca} since
it seems like one is probing the change in a quantum state (the
condensate) which is interacting with a thermal bath. The entropy
changes observed must be because the energy (temperature) stored in
the cloud undergoes a change due to addition of particles in the
condensate.


At a given energy where black hole solutions coexist with solutions
with condensates, the solutions with a condensate have higher entropy
(see Fig: \ref{Fig-SE1}). The process of condensation involves moving
charge from the black hole into the scalar cloud surrounding it. This
leads to an increase in entropy because the entropy of the black holes
increase as the density decreases at fixed energy.

In this process, since $\d E=0$,
$$0=\delta E \geq \delta \rho_{cond} E_0+\delta \rho_{T}E_1= \delta
\rho_{tot}E_0+\delta \rho_{T}(E_1-E_0)$$ with $E_1>E_0$. Here $E_1$
is the energy of the first excited state (in the single particle
language) from which particles are removed (which is likely the
highest occupied state). From this we get, $|\frac{\delta
  \rho_{cond}}{\delta \rho_{T}}|\geq \frac{E_1}{E_0}$ \, ($\delta
\rho_{T}<0$) which implies that if the (absolute value) of the slope
of the graph of $ \rho_{cond}\, {\rm vs.}\, \rho_{T}$ is greater than
unity (at $T=0$), this indicates a gapped system.

\begin{figure}[H]
  \centering \includegraphics[height=6cm]{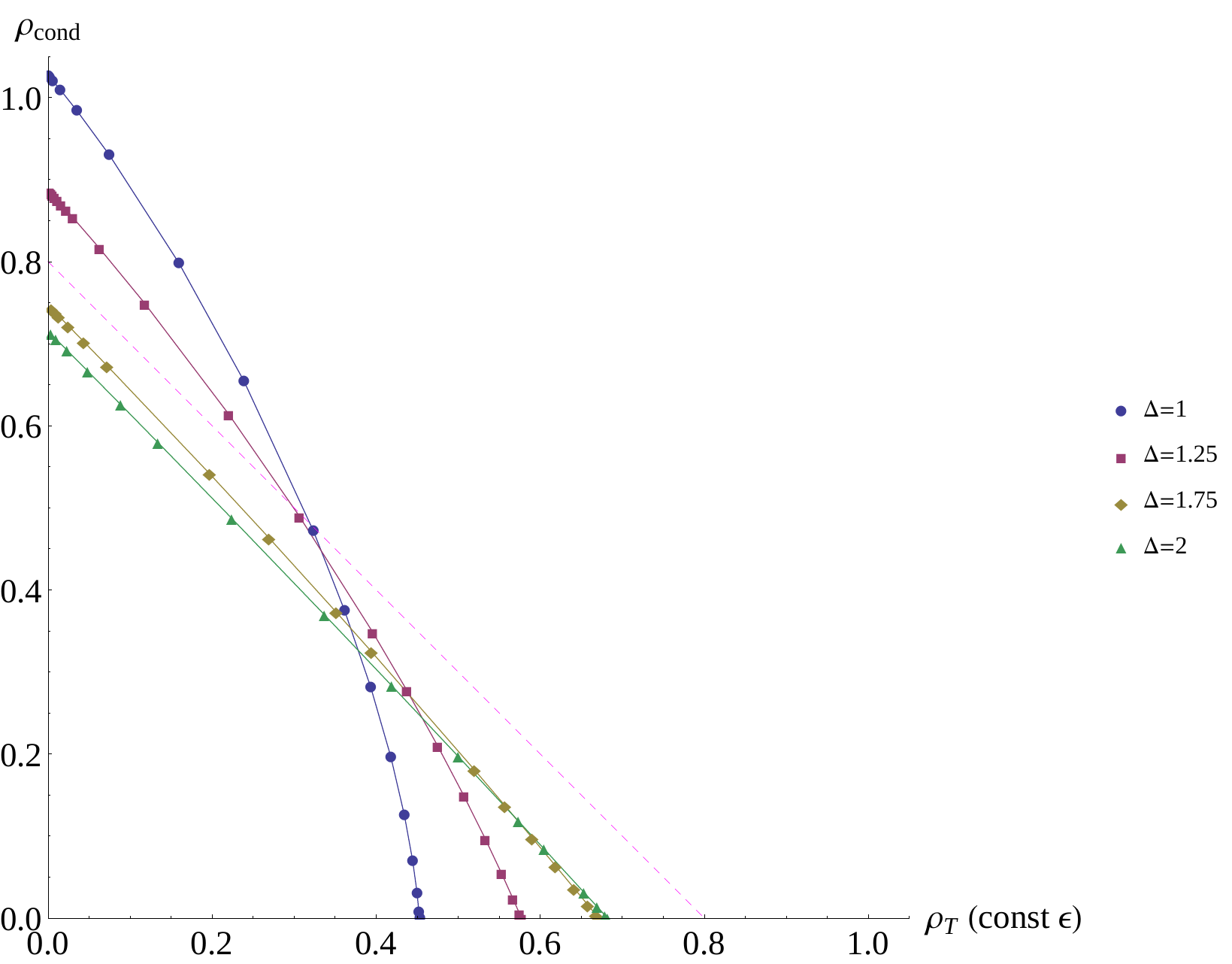}
  \caption{Occupancies of the condensate vs. excited states at fixed
    energy E}
  \label{qqe}
\end{figure}

These graphs for various scaling dimensions are shown in
Fig: \ref{qqe} along with a dashed (red) line line of unit slope. We
immediately observe that the near zero temperature values of
$\rho_{cond}$ (y-intercepts) are greater than the corresponding
x-intercepts which represent the initial population in the thermally
excited states. From a visual comparison, it is seen that the slope of
all curves is greater than unity (in the lower part of the curve near
$T_c$). However, we note that near the y-axis, the slope is nearly
unity at $T=0$ indicating the absence of an energy gap.

Note that at zero temperature, all the electric charge resides on the
condensate (and there is no black hole in the bulk).
\cite{Gubser:2008pf}-\cite{Horowitz:2009ij}. Thus, all the graphs
intersect the y-axis at nonzero values of $\rho_{cond}$. For a given energy,
the amount of $\rho_{cond}$ decreases with the scaling dimension. This is as
expected since the higher scaling dimension operators create states of
higher energy in the field theory. On the other hand, the onset of
condensation is represented by the x-intercept which shows a
systematic increase with the scaling dimension of the operator that
undergoes condensation also as expected. Fixing the total energy, we
can note that for smaller $\D$, the number of particles occupying the
condensate at $T=0$, is nearly double that in the thermal gas at
$T=T_c$ in contrast to the larger $\D$. This suggests that most of the
particles in the larger $\D$ systems are near the ground state (so
that energy changes only by little when these condense).

\section{Grand Canonical Ensemble: Chemical potential and equation of state}

\begin{figure}[H]
  \centering
  \includegraphics[height=6cm]{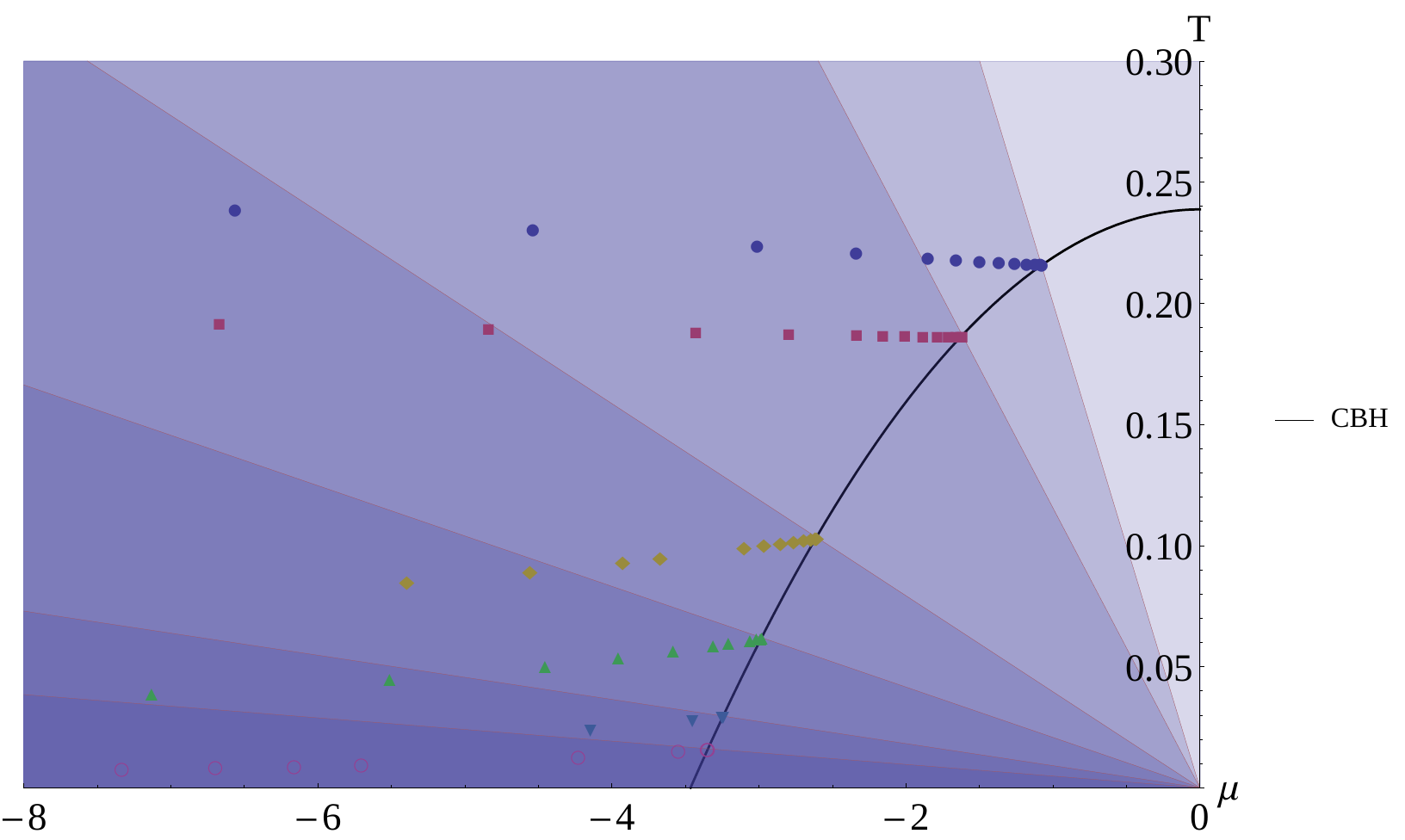}
\caption{\label{muTs} $\mu$ vs T for $s=\frac{1}{4\pi}$}
\end{figure}

In the grand canonical ensemble, thermodynamic states are labeled by
the chemical potential and temperature. Therefore, to begin with, we
try to understand how the various solutions cover the quadrant
$T>0,\mu<0$. In Fig: \ref{muTs}, the radial lines represent families
of solutions obtained by the scaling transformation, eqn:
\ref{scaling}, starting from any one of the solutions (represented by
colored dots) on that ray. We obtain straight lines because both $\mu$
and $T$ having dimensions of energy, transform in the same way under
scaling.
The solid black curve represents charged black holes without a
condensate and dots represent solutions with a condensate - all these
points have entropy density $s=1$. The various radial rays are
critical lines to the right of which there are NO solutions with a
condensate of that scaling dimension. For instance, in the region to
the right of the right most ray - there are no solutions with
condensates for scalar fields with scaling dimensions $\Delta>1$. The
blue dots representing the $\D=1$ condensates make their first
appearance on this ray; In this region, there are only black holes
without condensates (for the $\D=1$ operator) - there could be
operators of lower scaling dimension that could also condense. To the
right of the second most ray there are no solutions for scaling
dimensions $\Delta>5/4$ etc. As the scaling dimension increases, the
condensate solutions appear at larger (negative) values of the
chemical potential.

In the same figure, we see that, for fixed $\mu$ and $\D$, a
condensate solution always appears at low enough temperature (this
occurs at the intersection of the constant $\mu$ line with the ray
corresponding to the particular $\D$ chosen). The temperature at
which condensate solutions can be found decreases for large scaling
dimension, while for small scaling dimension, this critical
temperature actually increases. Note that in this figure, entropy is
being kept constant.

Thus, we see that for most of the quadrant the CBH solutions co-exist
with the condensate solutions. In this region, the question of which
phase dominates is then decided by the system having larger pressure
(lower grand potential). This is quite different from the
micro-canonical ensemble where the coexistence region is very small.

From the perspective of the microcanonical partition function
$$ \frac{\mu}{T}=-\log\frac{\Omega(E,V,N+1)}{\Omega(E,V,N)} $$ Thus,
$\mu$ represents the cost of adding a particle to the system while
keeping the thermodynamic internal energy $E$ constant. We may
interpret the higher energy cost for higher scaling dimension
condensates as being due to these being composite operators in terms
of single particle states.

\begin{figure}[H]
  \centering
\includegraphics[height=6cm]{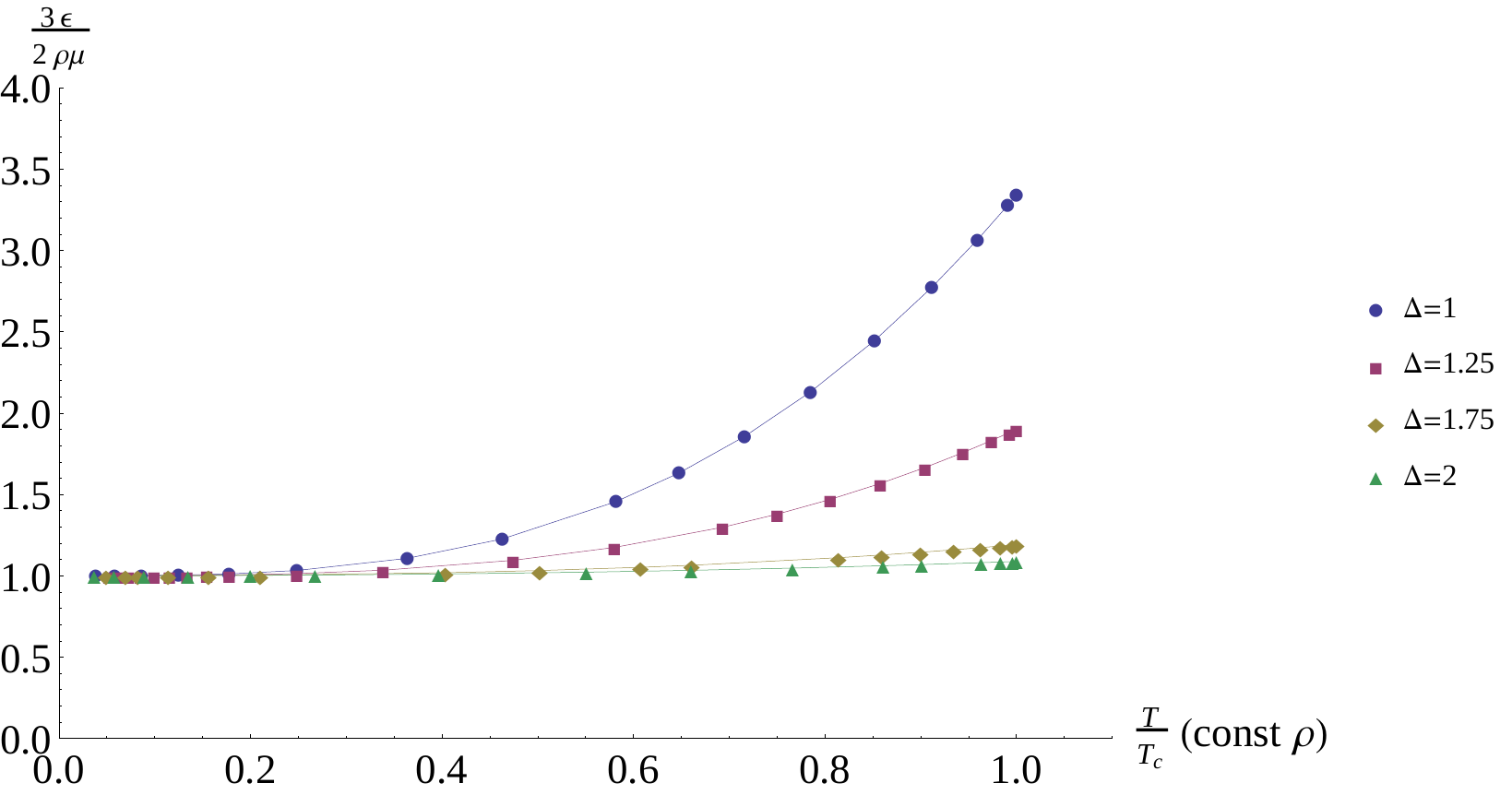}
\caption{$\frac{3\e}{2\mu\rho}$ vs T at constant $\rho$}
\label{emuratio} 
\end{figure}

At fixed {\em number density} after the onset of condensation, in
sharp contrast to the bare black hole, it is seen that the chemical
potential hardly changes. In any case, at $T=0$, we may expect that
the distinction between chemical potential and the ground state energy
disappears. This is indeed the case as shown by the graph in Fig:
\ref{emuratio} where the ratio of the energy per particle to the
chemical potential is plotted as a function of the temperature.
Note that at zero temperature, since the entropy goes to zero, because
of extensivity and the equation of state, the ratio
$\frac{\e}{\rho\mu}\to \frac{d-1}{d}$ - which is an artifact of the
conformal symmetry.



\subsection{Equation of state}

As mentioned earlier, a consequence of conformal invariance is the
tracelessness of the stress tensor of the field theory which leads to
the equation of state $\e=(d-1)p$. In the grand canonical ensemble, we
have another equation of state relating $\mu$ with its thermodynamic
conjugate, the number density $\rho$ (for fixed $T$ or $\e$).

\begin{figure}[H]
\begin{minipage}{.47\textwidth}
\flushleft
  \includegraphics[width=\linewidth]{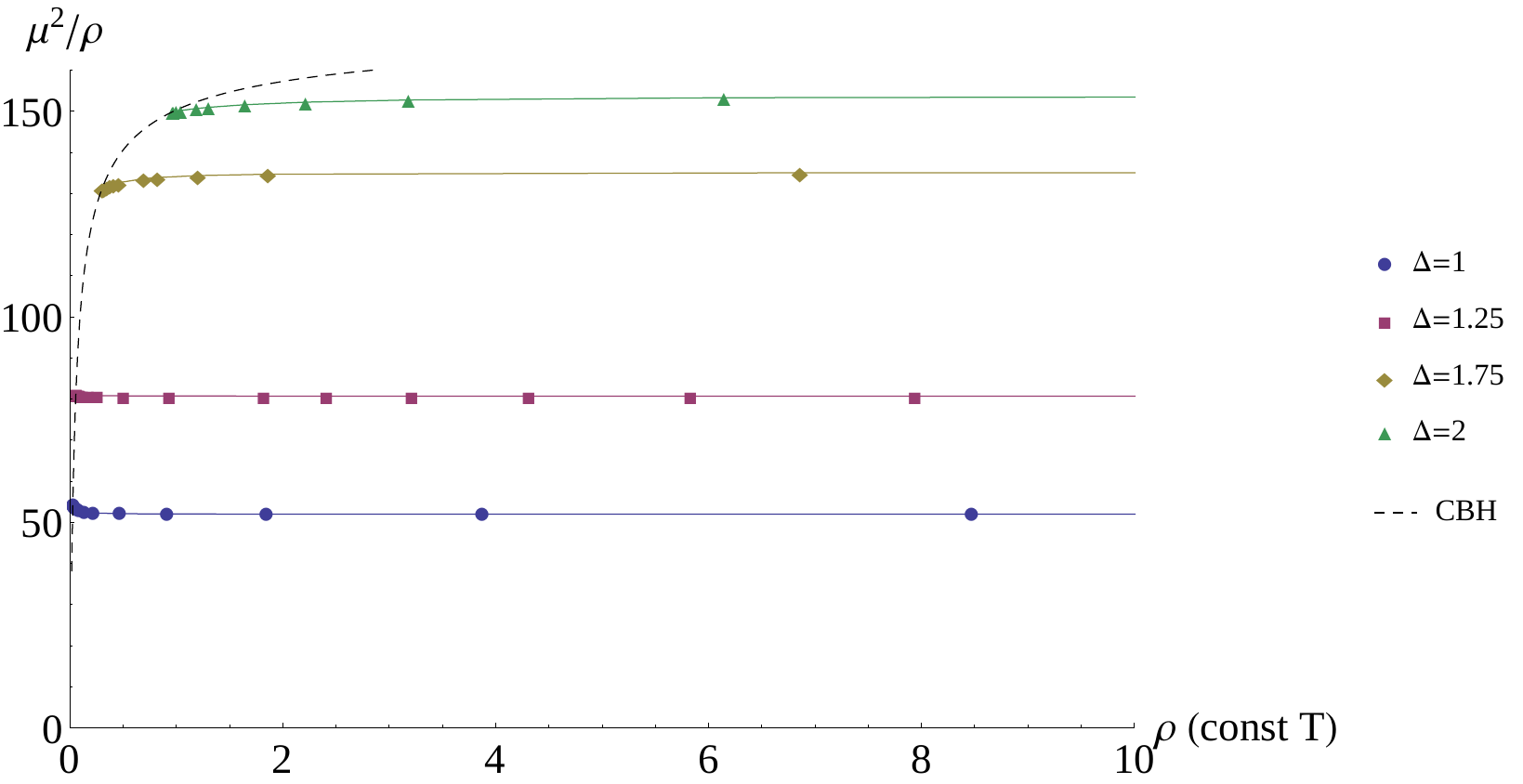}
\caption{$\frac{\mu^2}{\rho}$ vs $\rho$ at constant T}
\label{qmuT} 
\end{minipage}
\begin{minipage}{.47\textwidth}
\flushright
  \includegraphics[width=\linewidth]{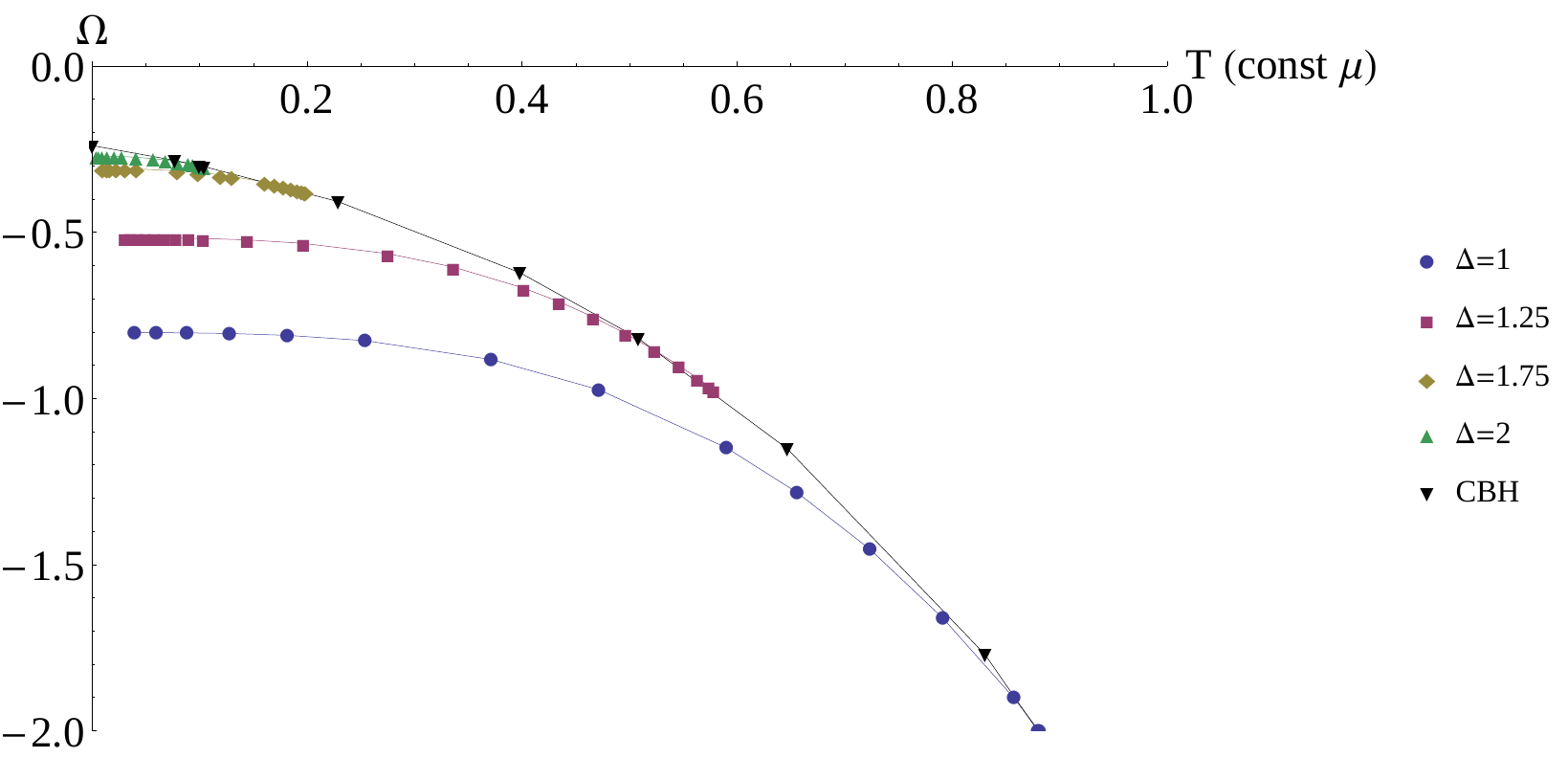}
\caption{$\Omega=\mu\rho=-p$ vs T at constant $\mu$}
\label{pressure} 
\end{minipage}
\end{figure}

Using extensivity, we can see that the equation of state takes the
form $\mu^2 =\rho\, f(\frac{T^2}{\rho},\Delta)$ (in $d=3$ again assuming
analyticity in $T$). The dependence on temperature however must be
determined numerically. Fig: \ref{qmuT} shows that in fact,
$\mu^2/\rho$ is nearly independent of $\rho$ when there is a
condensate (the dashed curves represents the equation of state
obtained from a charged black hole) The constant seems to depend
mildly on the scaling dimension with larger scaling dimensions having
more negative chemical potential values.


The phase diagram of the Einstein-Maxwell system alone was studied a
long time ago (for spherical boundary topology) \cite{Chamblin} where
it was observed that the charged black holes do not represent the low
temperature phases. The solutions with a condensate fill out the low
temperature region as seen in the graph of the grand potential as a
function of the temperature (for fixed chemical potential). Similar to
the canonical ensemble, the solutions with a condensate represent the
stable phase with a lower grand potential compared to the
corresponding black hole. As before, the phase transition is
continuous.

\section{Dependence on the Scaling dimension}
\label{delta}
In this section, we shall focus on the dependence of various
thermodynamic properties on $\D_-$ defined as the scaling dimension of
the operator that undergoes condensation.

\subsection{Power laws and Thermostatics}

The shape of the entropy curves (Fig: \ref{Fig-SE1}, Fig: \ref{sTQ})
suggests power law dependencies on the independent variable. Thus we
attempt to fit $s=s_0(\e-\e_0)^\alpha$ to our numerics, where the
constants $s_0,\e_0$ are functions of both $\D$ and $q$ as well as the
density $\r$ (which is being kept constant).

Such a fitting form is also suggested by experiment (for instance,
\cite{Luo:PRL}) where it is found that the entropy for ${\rm Li}^6$
atoms in a trap could be fitted with the formula
$s=s_0\left(\frac{\e-\e_0}{\e_F}\right)^\a$ with an exponent $\a=0.6$.
Their system thus behaved like a degenerate Fermi gas and $\e_F$ was
interpreted as the Fermi energy while $\e_0$ is the ground state
energy per particle.

The comparison with the holographic system must be made with
circumspection: the holographic system is both conformal and
relativistic - although both these symmetries are broken by the
temperature and chemical potential. The dual boundary system lives in
2 noncompact dimensions whereas the above experiment is in a trap in
3D. Nevertheless, the interpretation of $\e_F$ is likely to be
profitable.

From the expression for entropy, we can obtain the dependence of the
energy on temperature by inverting the equation
$\frac{1}{T}=\frac{\del s}{\del \e}$. Thus, $\e=\e_0+(\frac{\e_F
  ^\a}{\a s_0})^{\frac{1}{\a-1}}\, T^{\frac{1}{1-\a}}$ and hence the
exponent for the specific heat is determined $c_v\sim
T^{\frac{\a}{1-\a}}$. 

\begin{minipage}{.47\textwidth}
  \flushleft
\begin{tabular}{|c|c|c|c|}
\hline
$\D$ & $ s_0$ & $\e_0$&$\a$ \\\hline
1   & 2.28 & 0.43 & 0.68 \\ \hline
5/4 & 2.53 & 0.53 & 0.67 \\ \hline
7/4 & 3.51 & 0.69 & 0.61 \\ \hline
2   & 4.43 & 0.73 & 0.57 \\ \hline
9/4 & 5.80 & 0.76 & 0.52 \\ \hline
12/5& 8.0  & 0.77 & 0.52 \\ \hline
\end{tabular}
\label{TablesE}
\end{minipage}
\begin{minipage}{.47\textwidth}
\begin{figure}[H]
\flushleft
  \includegraphics[width=\linewidth]{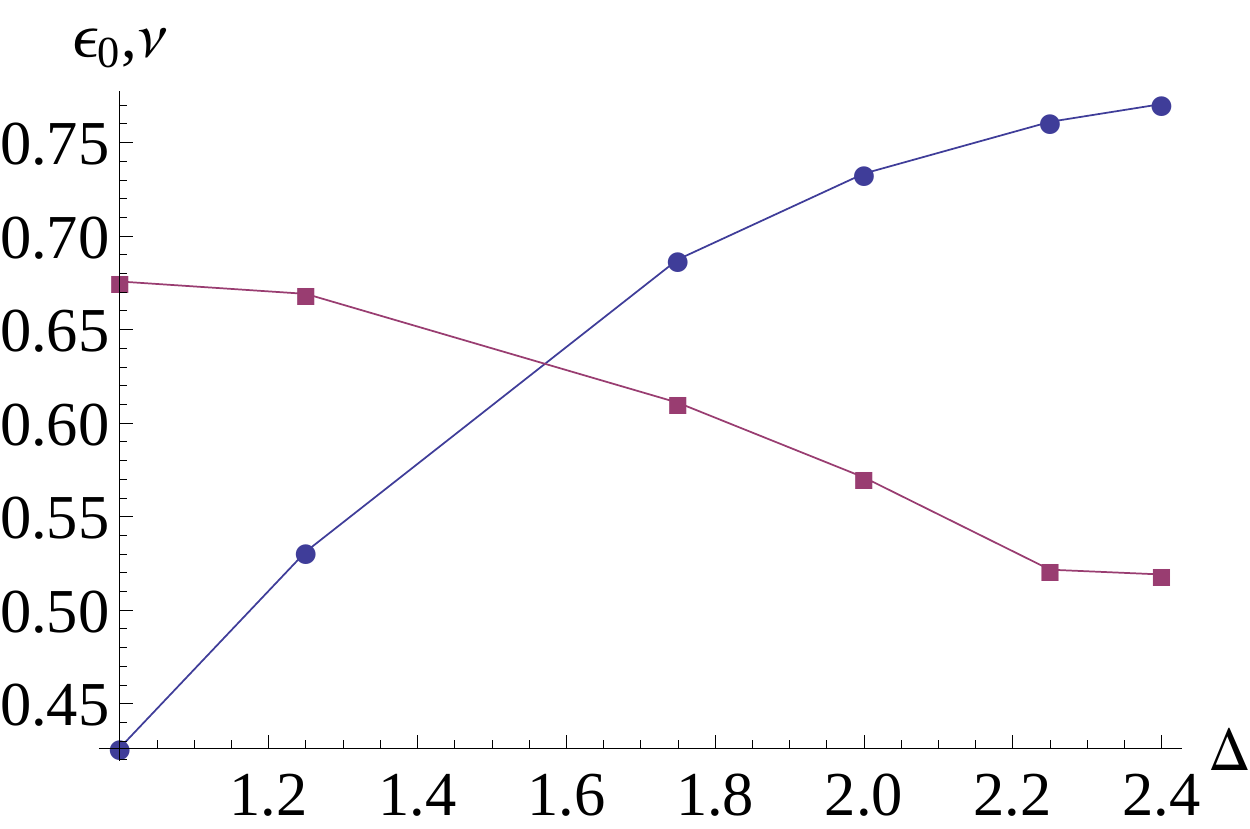}
  \caption{$\e_0$ and $\n$ vs $\D$}
\label{seFit} 
\end{figure}
\end{minipage}
\vspace{0.25cm}

It is noteworthy that we obtain an excellent fit to the numerical
data with the above fitting form. The best fit values for the
parameters of the fit are shown in Fig: \ref{seFit} and the
Table: \ref{TablesE} above. The graph shows that the exponent varies
from nearly $\a=2/3$ for $\D=1$ to $\a=1/2$ for large $\D$. Note also
that the energy per particle increases with the scaling dimension, but
seems to be saturating for large $\D$. 

For the dependence of the entropy on the density (keeping the energy
held constant), the expression
$s=s_0\left(\frac{\rho_0-\rho}{\rho_0-\rho_c}\right)^\beta$ fits the
data near the onset of condensation $\rho_c$ (but deviates somewhat
near $T=0$ for the smaller scaling dimensions). In the table
\ref{srtable}, we show the best values of $\rho_0$ and $\beta$ as well
as the values of the critical density $\r_c$ (note that both
$\rho_{0,c}$ depend on the constant energy).  The accompanying graph
shows the ratio of $\frac{\rho_0}{\rho_c}$ which illustrates the large
variation with the scaling dimension.

\begin{minipage}{.47\textwidth}
  \flushleft
\begin{tabular}{|c|c|c|c|}
\hline
$\D$ & $ \frac{\r_0}{\rho_c}$ & $\r_c$&$\b$ \\\hline
1   & 2.26 & 0.45 & 0.64 \\ \hline
5/4 & 1.54 & 0.56 & 0.64 \\ \hline
7/4 & 1.113 & 0.67 & 0.60 \\ \hline
2   & 1.05 & 0.68 & 0.57 \\ \hline
9/4 & 1.02 & 0.68 & 0.51 \\ \hline
12/5& 1.01 & 0.69 & 0.52 \\ \hline
\end{tabular}
\label{srtable}
\end{minipage}
\begin{minipage}{.47\textwidth}
\begin{figure}[H]
\flushleft
  \includegraphics[width=\linewidth]{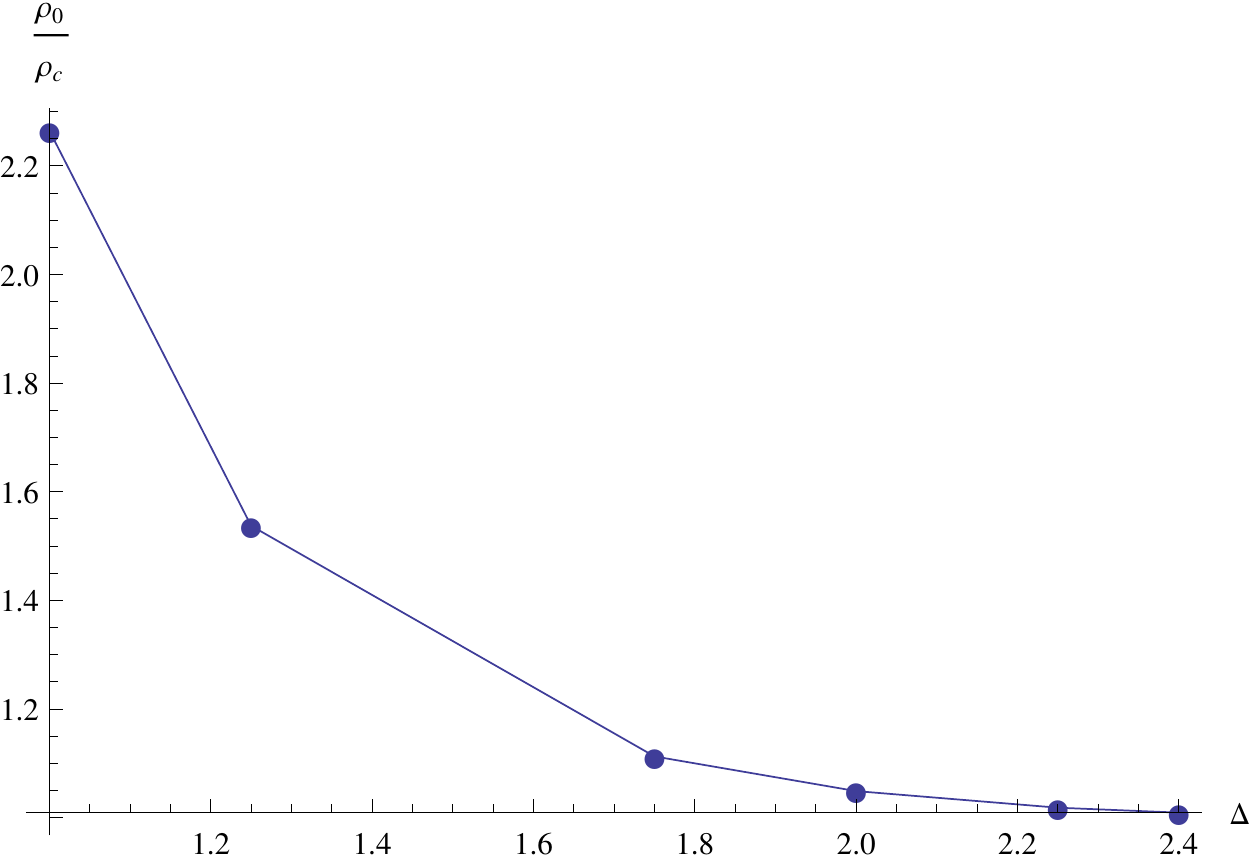}
  \caption{$\frac{\rho_0}{\rho_c}$ vs $\D$ at fixed $\e$}
\label{rrhoc} 
\end{figure}
\end{minipage}
\vspace{0.25cm}
The exponents in the two tables above agree reasonably well.


We can read off the ground state energy per particle from the
y-intercept in the graph Fig: \ref{qqe}, or equivalently from
$\e_0=\frac{\e}{\rho_0}$ where $\e$ is the constant energy at which
Table: \ref{srtable} was obtained. This can be compared with the zero
temperature value as mentioned in Table: \ref{Table} (keeping in mind
that the different density used for the latter table). These different
evaluations agree to numerical accuracy (as they must).

In the canonical ensemble, we can numerically fit a polynomial to the
$\e$ vs $T$ points and from this obtain a graph of the specific heat
itself - this was shown in Fig: \ref{ecv}. The specific heat behaves
as $C_v\sim T^\s$ as shown in the table below - where one can see that
the ground state energy $\e(T=0)$ scales (nearly) linearly with the
scaling dimension. These values for the ground state energy and
exponents agree with those obtained from the $s$ vs. $\e $ curves (as
they must) $\s=\frac{\a}{1-\a}$.

\begin{center}
\begin{tabular}{|c|c|c|c|c|c|c|c|}
\hline
$\D$ & $\e$ & $\tilde\e_0$&$\e_F$&$
\frac{\rho_0}{\rho_c}-1$&$\mathcal{O}_0$&$\s$ \\\hline
1  & $0.15+3.6 T^{\s+1}$ & 0.19 & 0.29 & 1.265  & 4.7 & 2.06\\\hline
5/4& $0.19+4.8 T^{\s+1}$ & 0.22 & 0.25 & 0.552  & 5.6 & 2.03\\\hline
7/4& $0.24+14.5 T^{\s+1}$& 0.27 & 0.143& 0.113 & 21  & 1.6\\\hline
2  & $0.26+37 T^{\s+1}$  & 0.29 & 0.09 & 0.049 & 67  & 1.3\\\hline
\end{tabular}
\label{Table}
\end{center}

We also showed the dependence of the order parameter on the density
(keeping $\e$ fixed) in Fig: \ref{oqe}. The vev is well fitted by
curves of the form $\langle \mathcal{O} \rangle=
\mathcal{O}_0\sqrt{\frac{\rho}{\r_c}-1}$ where the constant
$\mathcal{O}_0$ is shown in Table: \ref{Table}. The order of
magnitude difference in the prefactor $\mathcal{O}_0$ (Table: 
\ref{Table}) between the lower scaling dimensions and the higher ones
is noteworthy - a similar such difference was pointed out as defining
either sides of the BEC-BCS crossover in \cite{Stringari}. Such a
large difference was also observed in the earlier work \cite{KKY} in
the condensate fraction in the core of vortices and dark solitons.

\subsection{Onset of Condensation}
We plot the value of $T_c$ at which condensation starts as a
function of scaling dimension while keeping the chemical potential
fixed. 
For a given $\D$, as $q$ increases the onset of condensation $T_c$
also increases for fixed $\mu$. For larger $\D$, the the temperature
of onset decreases as expected.
\begin{figure}[H]
  \centering
  \includegraphics[width=0.6\textwidth]{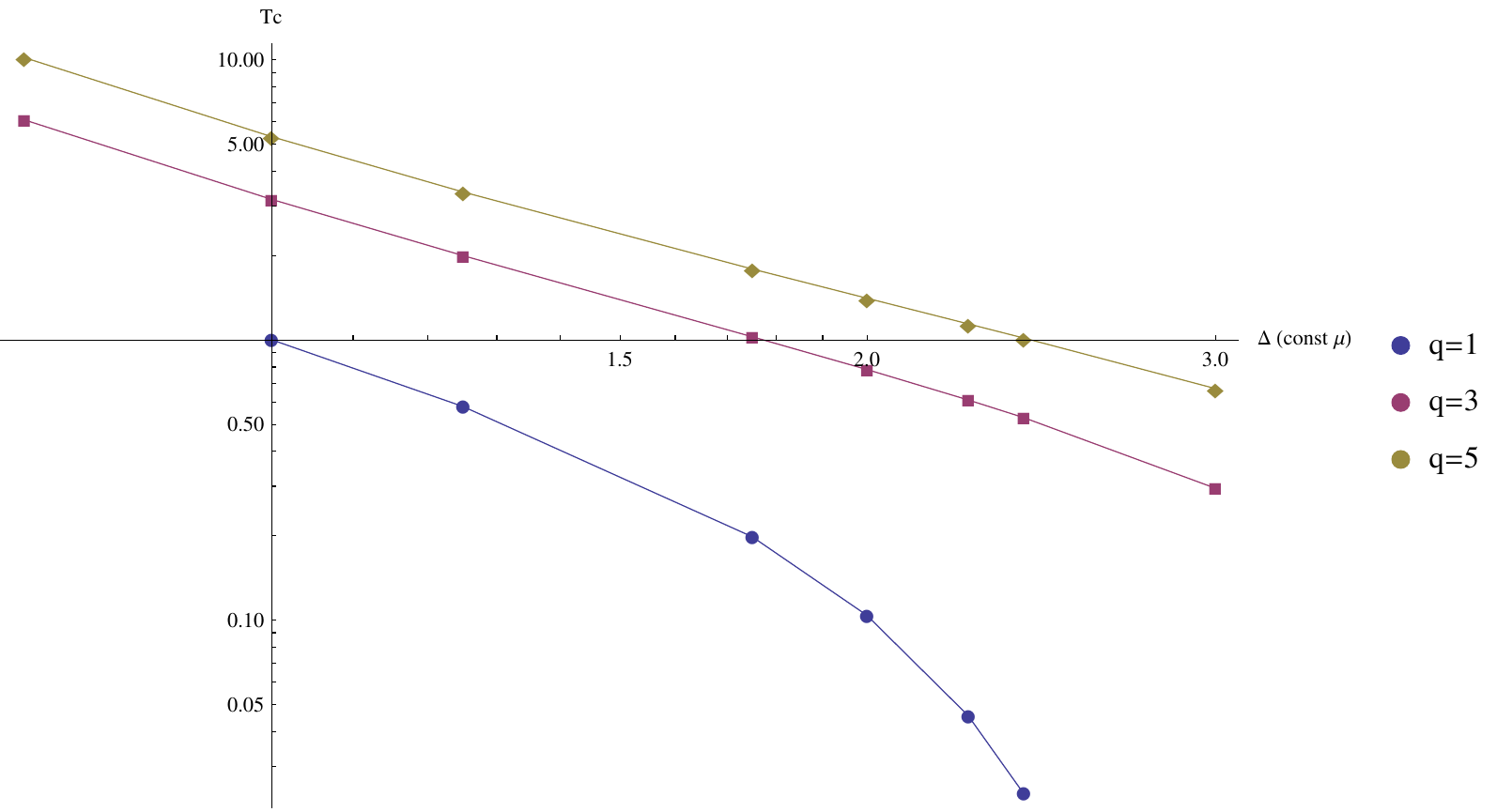}
  \caption{$T_c$ vs. $\Delta$ at constant chemical potential}
  \label{critT}
\end{figure}
Our proposal is to compare this graph with computations of the
variation of the condensation temperature with the interaction
strength as in for instance cold atom systems \cite{what} or neutrons
\cite{Sunethra}. In these systems, the possibility of BEC-BCS
crossover leads to the presence of a new energy scale, namely the
Fermi energy (apart from the healing length). Such a possibility was
already explored earlier in \cite{KKY} where the core densities of
solitons were argued to show signs of a Fermi energy.

On dimensional grounds, the appearance of a new length scale might be
expected to alter the functional dependence of $T_c$ on $\D$. In the
graph of $T_c$ vs $\D$, which is shown on logscale, we can see that
for large $q$ where the back-reaction of the scalar cloud is
insignificant, the $T_c$ is a power of $\D$.

But, for $q=1$, where the back-reaction plays a significant role, we
might want to see a difference arising for $\D>=2$. Such a change in
character maybe attributed to a new scale either arising or
disappearing - but the graph is not conclusive.


\subsection{Joule Thomson Coefficient}

From Fig: \ref{muTs}, where one sees two different slopes for the
chemical potential graphs, one might be tempted to infer a qualitative
difference. This difference is clearly visible in a graph of how the
density changes while adiabatically (keeping entropy constant)
changing the temperature - i.e., the coefficient of volume expansion.
The $\D=1,1.25$ curves show a monotonic increase in the density at
fixed entropy whereas the $\D=2,2.25$ show a decrease. For
$\Delta<\frac{d}{2}$ the chemical potential decreases (becomes more
negative) as the temperature decreases (with entropy fixed), whereas
for $\Delta<\frac{d}{2}$, the chemical potential increases (the
solutions for $\Delta=0.75$ are numerically inaccurate). Contrast
these with the behaviour of the chemical potential for the charged
black holes without any condensate - which is the flat horizontal
curve towards the bottom of the graph.

From this one can obtain the Joule Thomson coefficient
$\mu_{JT}=\frac{1}{\r}\frac{d\rho}{dT}|_s$. In a graph of the density
against temperature at constant entropy, the regions where the slope
is negative is the region where cooling occurs upon expansion.
\begin{figure}[H]
\flushleft
\begin{minipage}{.5\textwidth}
  \centering
  \includegraphics[width=\linewidth]{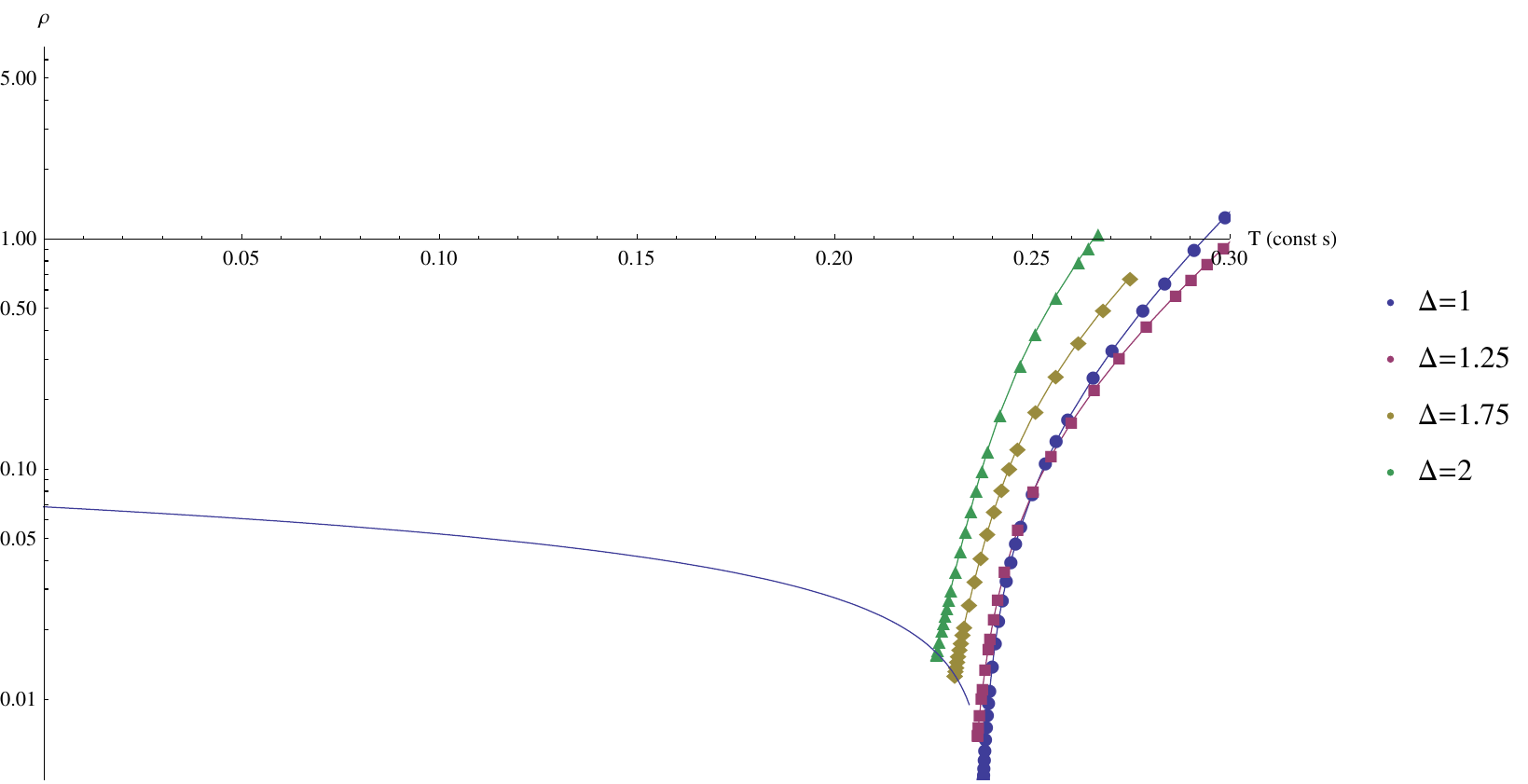}
  \captionof{figure}{$\rho$ vs $T$ for constant $s$ (q=1)}
  \label{JT}
  \end{minipage}%
\begin{minipage}{.5\textwidth}
  \flushright
   \includegraphics[width=\linewidth]{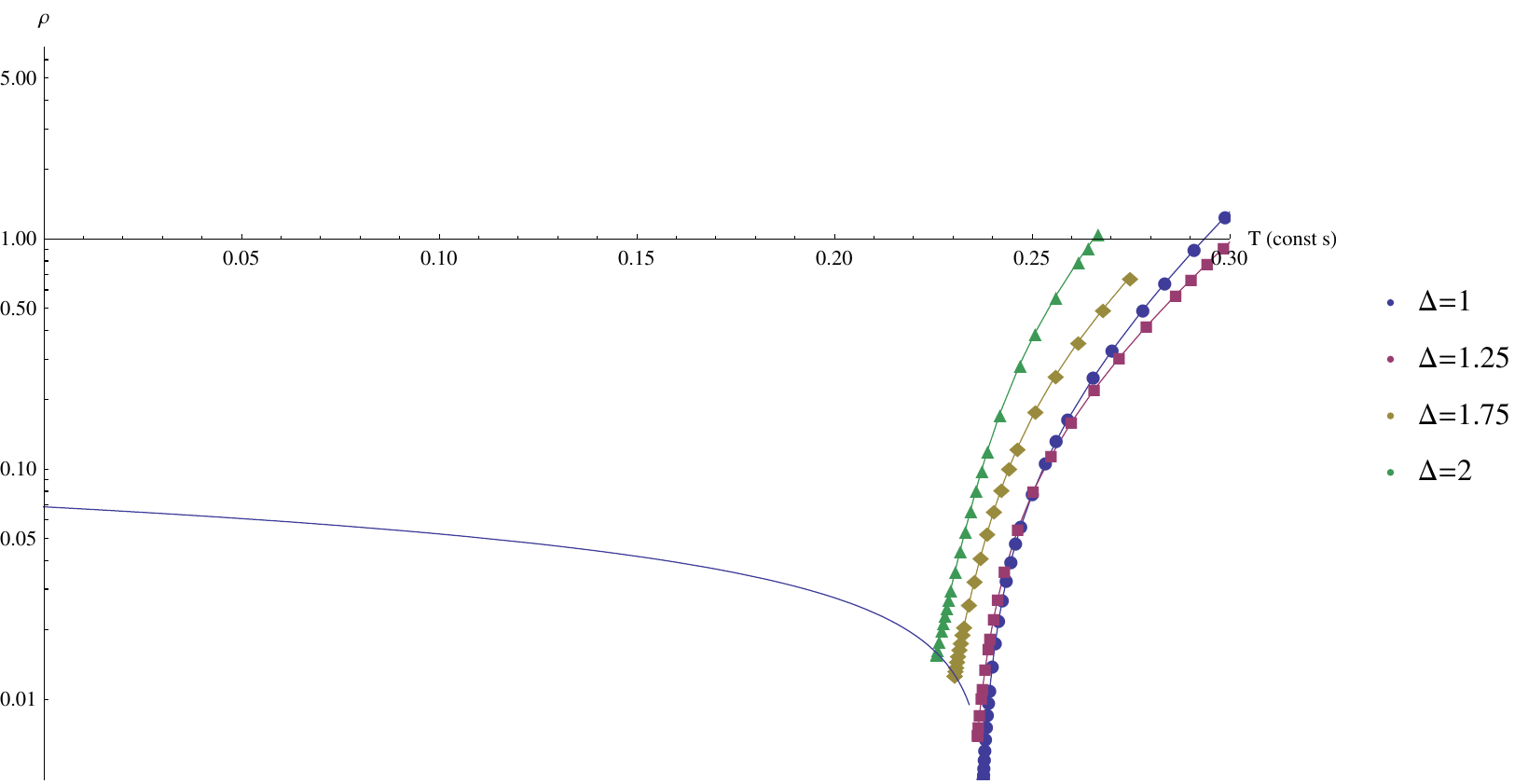}
  \captionof{figure}{$\rho$ vs $T$ for constant $s$ (q=5)}
  \label{jt1}
\end{minipage}
  \end{figure}
As seen in the graph, for small $q\sim 1$, the large $\D$ superfluid
undergoes cooling upon expansion. Whereas, for either small $\D$ or
for large $q$ (and any $\D$), expansion is accompanied by temperature
increase. The horizontal curve is the expansion coefficient of the
black hole. This curve terminates at the rightmost point.

\begin{figure}[H]
\flushleft
\begin{minipage}{0.45\textwidth}
  \centering
  \includegraphics[width=\linewidth]{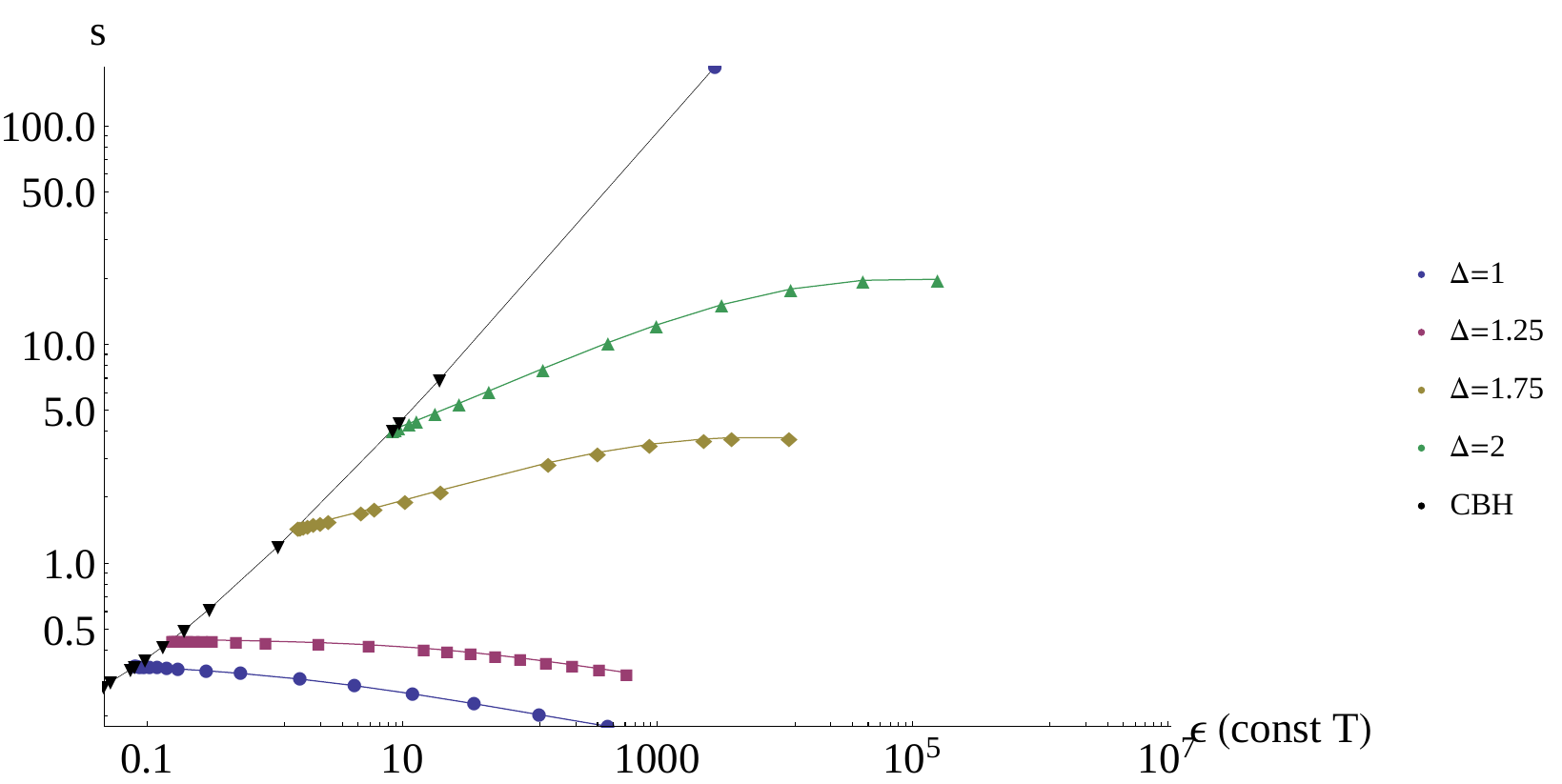}
  \caption{Entropy vs. Energy constant T}
  \label{esT}
\end{minipage}
  \begin{minipage}{0.45\textwidth}
    \flushright
  \includegraphics[width=\linewidth]{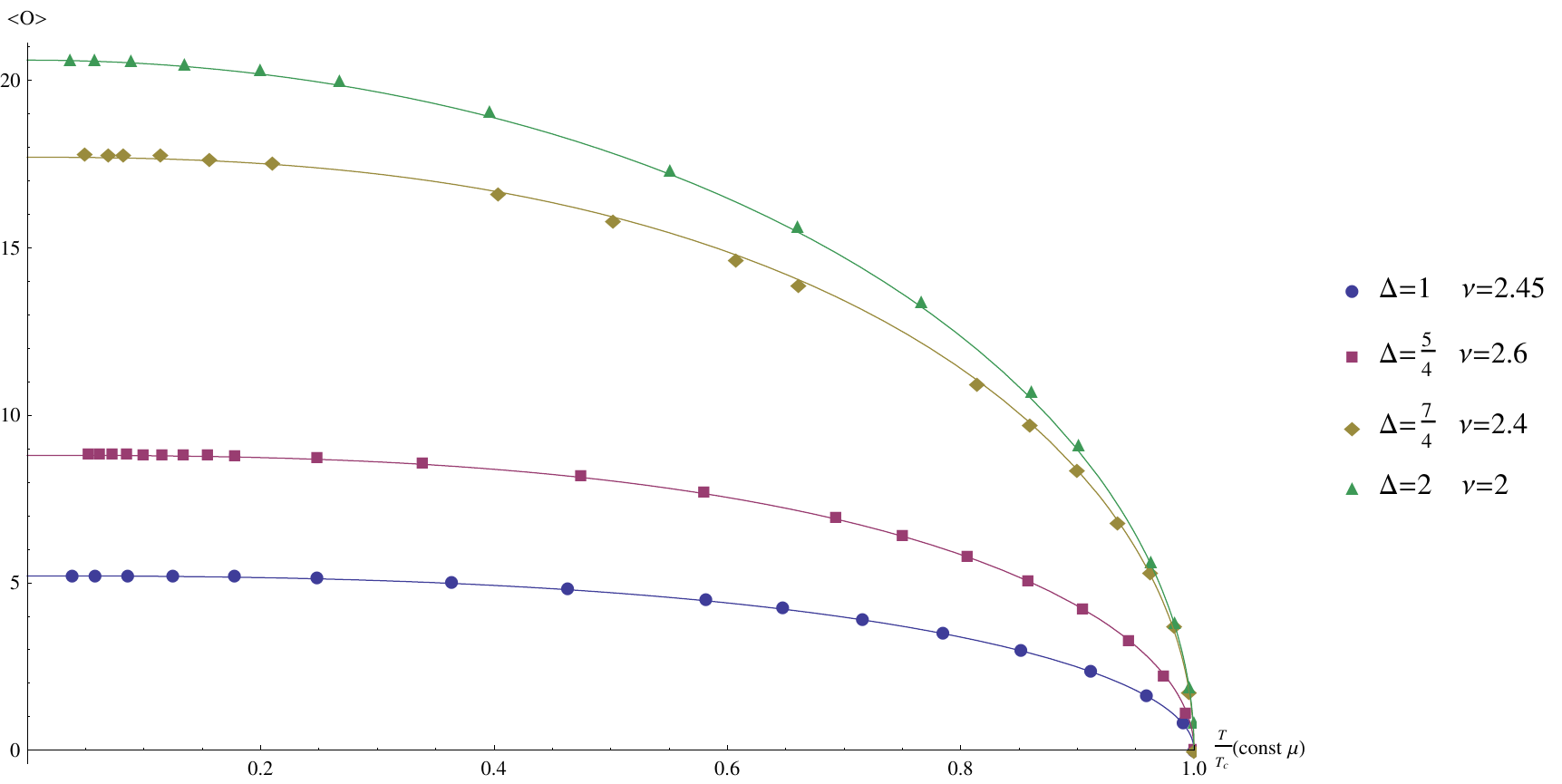}
\caption{\label{vvevT} $\frac{<{\mathcal O}>}{{\mathcal O_0}}$ vs
  $\frac{T}{T_c}$ at constant density $\rho$ for $q=1$}
\end{minipage}
\end{figure}

On the other hand, consider the slope $\left(\frac{\del s}{\del
  \e}\right)_T=\frac{\left(\frac{\del s}{\del
    \rho}\right)_T}{\left(\frac{\del \e}{\del \r}\right)_T} $. Since
the temperature is fixed, the property that is being varied is the
number of particles. From the graph (Fig: \ref{esT}), it is clear that
entropy decreases for the $\Delta<\frac{d}{2}$ by adding particles.
This is possible only if there was an attraction between the particles
- because the decrease in entropy can only mean a decrease in phase
space volume which is possible if the particle motions are somehow
restricted.
In contrast, the higher scaling dimensions show an initial repulsion
which would lead to the added particles occupying a larger phase space
volume leading to an increase in entropy. Note that eventually (at
large densities) we expect that all additional particles occupy the
ground state and so the entropy will end up at some nonzero value due
to the constant (nonzero) temperature.

\subsection{Solution profiles \label{soln-profile}}
\begin{figure}[H]
\flushleft
\begin{minipage}{0.45\textwidth}
  \centering
  \includegraphics[width=\linewidth]{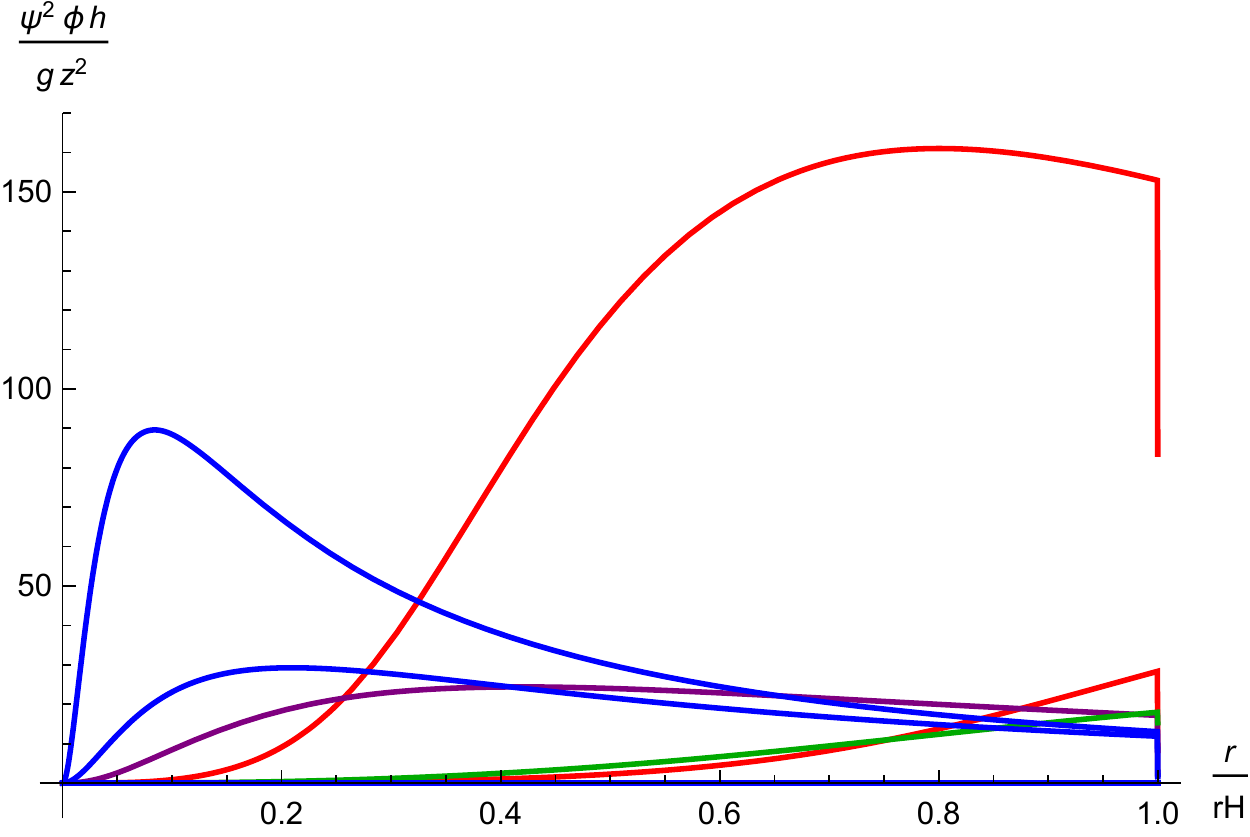}
  \caption{Charge density along the radial direction}
  \label{charged}
\end{minipage}
  \begin{minipage}{0.45\textwidth}
    \flushright
  \includegraphics[width=\linewidth]{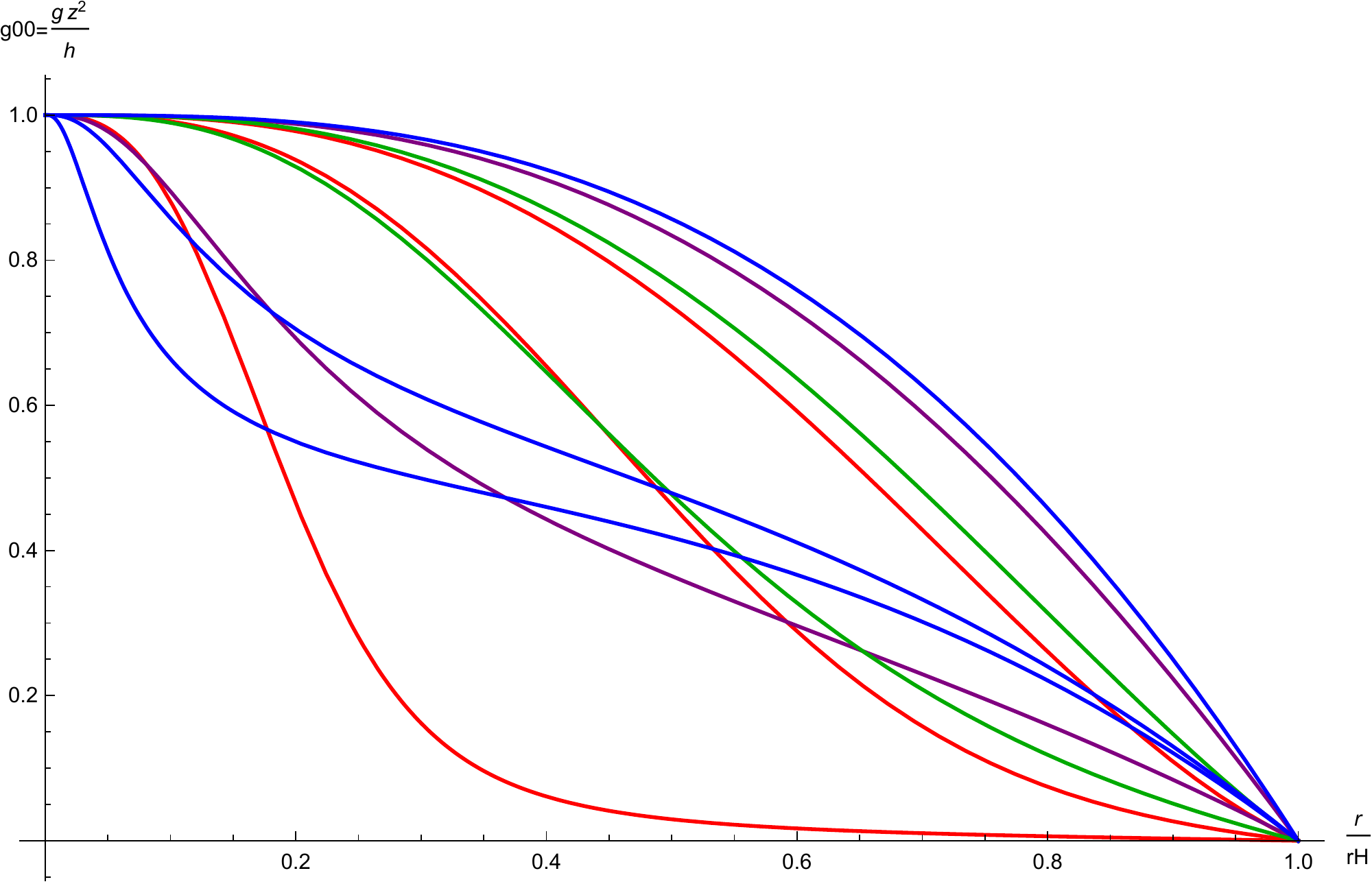}
\caption{\label{g00} $z^2 g_{00}$ along the radial direction}
\end{minipage}
\end{figure}

The features described above maybe traced back to the properties of
the gravitational side of the hologram by considering the bulk
equations \ref{eqn}. Asymptotic analysis tells us that $\psi\sim
z^{\D_-}$ \ref{asymp} near the boundary. Thus, the electric charge
density in the scalar field, expressed as $\rho(z)=\frac{\psi^2 A_0
  h}{g z^2 q^2}$ gets pushed away from the boundary as the scaling
dimension increases (Fig: \ref{charged}) because the electric scalar
potential in the bulk is essentially linear (and vanishes at the
horizon).
The behaviour of the bulk metric on the other hand, carries
information about the temperature. In this case, increasing the charge
in the scalar field near the horizon will decrease the curvature
$g'_{00}$ of the metric as we can see from the gravity equations in
\ref{eqn}. For small scaling dimensions on the other hand, any extra
charge is pushed out close to the boundary as seen in Fig: 
\ref{charged} - thus having small effect on the metric near the
horizon. For larger scaling dimensions, the extra charge localizes
near the horizon Fig: \ref{charged} and hence this leads to a much
larger decrease in slope leading to very low values for the
temperature. Note that this effect of the charge competes with the
increase in $g'_{00}$ due to the $d z$ term in \ref{eqn} which is
present even for the black hole without any scalar field.

In Fig: \ref{g00}, we see that the metric profile for small scaling
dimensions (the curves drawn in blue) is similar to the metric of the
charged black hole without a condensate (the outermost blue
curve). However, for large scaling dimensions and larger values of the
condensate, $g_{00}$ flattens out in the region near the horizon. For
each scaling dimension (as indicated in the legend), we have shown a
pair of curves - the outer curve having lower value of the
condensate. We also see that the even for moderate condensation, the
second derivative $g'' _{00}$ differs in sign for the larger scaling
dimensions (turns positive) compared to the smaller ones (remains
negative). Presumably, this is the reason behind the differing
properties of the large and small scaling dimensions.

This does not explain the subsequent increase in temperature for the
smaller scaling dimensions (Fig: \ref{JT}). Also, the temperature
increases with $\rho$ for large $q$ independent of the scaling
dimensions. Presumably, in this range, the temperature is not affected
much by the scalar cloud since the fraction of charge contained in the
cloud is small as compared to the black hole charge. Thus, the
behaviour is essentially that of the charged black hole.



Further, moving some electric charge from the black hole into the
scalar cloud keeping the total charge fixed leads to an initial
increase in energy (from the coulomb repulsion). But, eventually the
Coulomb repulsion cost attains a maximum since as far as the electric
charge is concerned, the scalar cloud and black hole are symmetric.
Thus, at fixed entropy, energy of the system decreases when electric
charge is transferred from the black hole to the scalar cloud. The
amount of decrease in energy is less Fig: \ref{Fig-SE1} for the smaller
scaling dimensions since, for these, the electric charge is more
delocalized in the bulk (note that in Fig: \ref{Fig-SE1}, the total
density is being varied).

Above, we attempted to explain the observed features of the system
entirely in the language of the bulk gravitational system. We might
expect that these ``bulk explanations'' can be translated entirely
into the language of the boundary theory. In this context, one might
expect that the interpretation of the radial direction as the
(renormalization) scale of the field theory could play a role as well.

\subsection{Order parameter}
The condensate order parameter - which is the vacuum expectation value
of the operator that breaks the $U(1)$ symmetry spontaneously, shows
the characteristic behaviour as demonstrated in other works- except
that we may notice a systematic increase in the zero temperature value
with the scaling dimension. The figure Fig: \ref{vvevT} also includes
$\sqrt{1-x^\nu}$ best-fit curves with exponents as indicated in the
plot legend and the order of magnitude difference between
$\D>\frac{3}{2}$ and $\D<\frac{3}{2}$ maybe noted. Another feature is
the trend as $\D$ increases. For $\D=1,5/4,7/4$ the exponent decreases
with increasing $\D$ whereas for $\D=2,9/4,12/5$ the exponent
increases (the data for $\D=9/4, 12/5$ are close enough to be fit with
either exponent, but in any case we recall that $\D=9/4$ data is
numerically inaccurate).

\subsection{Compressibility}
\begin{figure}[H]
\flushleft
\begin{minipage}{0.45\textwidth}
  \centering
  \includegraphics[width=\linewidth]{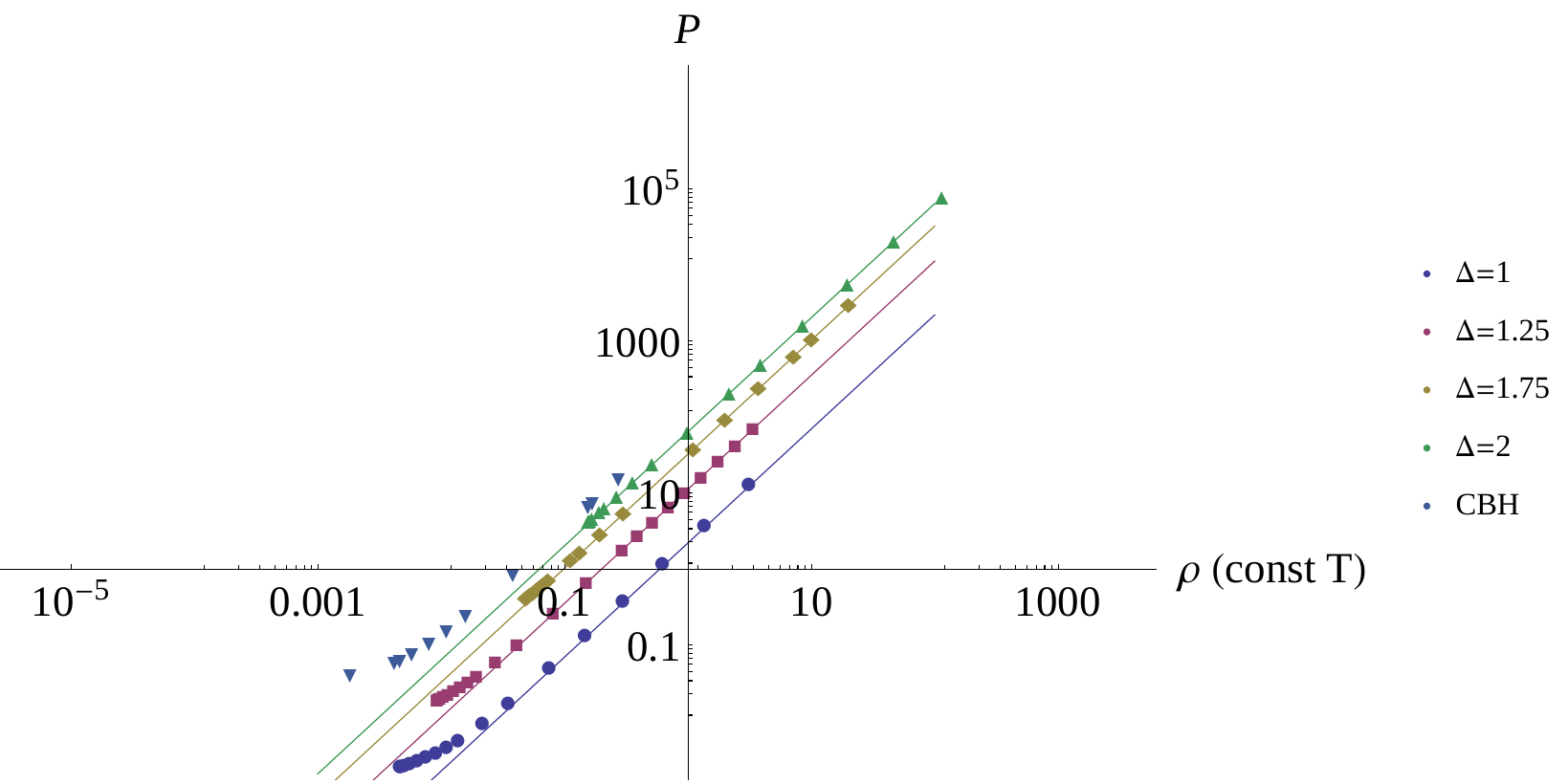}
  \caption{Pressure vs. Density constant T}
  \label{comp}
\end{minipage}
  \begin{minipage}{0.45\textwidth}
    \flushright
  \includegraphics[width=\linewidth]{vvevT.pdf}
\caption{\label{vevT} $\frac{<{\mathcal O}>}{{\mathcal O_0}}$ vs
  $\frac{T}{T_c}$ at constant density $\rho$ for $q=1$}
\end{minipage}
\end{figure}

Observing that the number density (Fig:  \ref{qqe}) or condensate (Fig: 
\ref{vevT}) at zero temperature is much larger for the smaller scaling
dimensions leads us to compare the compressibility of the various
cases. The larger occupancy suggests that the compressibility is
larger.

We can obtain both isothermal and adiabatic(isentropic)
compressibilities defined as
$$\k_{T,S}=-\frac{1}{\rho}\frac{d\r}{dP}$$ from the numerical
solution. The isothermal graph of pressure vs. density is shown in Fig:
\ref{comp} (note the log-scale) along with best fit curves of the
form $$ P=P_0\rho^\frac{3}{2}$$.

From these, we may extract the isothermal and adiabatic
compressibilities, and obtain the ratio
$\frac{\kappa_T}{\kappa_S}$. This ratio should be greater than unity -
in fact, one can define an effective number of degrees of freedom
$d_f$ by $\frac{\kappa_T}{\kappa_S}=1+\frac{2}{d_f}$. For all values
of $\D$, our numerical accuracy does not distinguish $\k_s$ and $\k_T$
and so the number of degrees of freedom is essentially infinity. For
this to be deemed sensible, we should be able see that $d_f$ tracks
finite N corrections as befits a holographic interpretation (perhaps, this
be done for charged black holes alone) .

A second feature that is noteworthy is the behaviour near the bottom
end of the graph (at low densities) which is near $T_c$. The
compressibilities are clearly becoming larger (pressure is becoming
constant) - perhaps it might be worth exploring if the compressibility
actually diverges.

\section{Dependence on q}

In this section, we shall study how the above features vary as we vary
the parameter $q=\frac{e\l}{\sqrt{G_N}}$. Since $G_N\sim 1/N^2$,
varying $q$ is interpreted as varying $N$ which is related to number
of colors in the gauge theory. The gauge coupling $\l$
could be interpreted in a variety of ways as discussed in the
introduction. We can also view $q$ as standing proxy for $e$ which
characterizes the strength of the interaction between particles in the
condensate and the thermal bath. Of course, we may simply take the
bulk gravitational point of view in which case $q$ stands in for
$G_N$.

Each thermodynamic aspect discussed in the previous sections has a
corresponding tale to tell as we vary $q$. We shall present only those
where the variation with $q$ is either significant or
interesting. Since $q$ appears as an overall factor multiplying the
matter part of the action, we may expect that thermodynamic quantities
involving only the scalar field and Maxwell field will not be
significantly affected by variation in $q$. Therefore, we focus on
thermodynamic properties involving the metric and a matter field such
as entropy vs. number density etc. At large $q$, the matter fields do
not affect the gravitational fields - in practice, we see that even
for $q\sim 5$, the metric profile is hardly changed by the
condensation of the scalar field at moderate values of the density and
away from $T=0$.


In each case, we consider the boundary terms at $z=0$ under an
arbitrary variation of the bulk fields. If we identify the boundary
chemical potential as $\mu=A_0(z=0)$, then the number density picks up
the $q-$factors $\rho=\frac{1}{q^2} A_0$ (via the functional
derivative $\frac{\delta \log Z} {\delta \mu} = \rho$). We may also
identify $\mu=\frac{1}{q} A_0(z=0)$ whence $\rho=\frac{1}{q} A_0$. In
the first case, $A_0$ maybe interpreted as the dual of a quark
current, since $\rho$ is proportional to $N^2$, while in the second
case, $A_0$ is more like ``baryon'' current. Similarly, the number $e$
controls the interpretation of the condensate - whether ``pairs''
($e=2$) or ``single'' ($e=1$) particles make up the condensate.

The parameter $\lambda$ has the interpretation of the interaction
strength between the particles (after all, it {\em is} the YM coupling
constant in the bulk). In the superfluid literature, this interaction
strength between the particles is tunable (usually with the aid of the
Feshbach resonance) and is captured by the scattering length
$a$. Unfortunately, all these cases are covered by the same equations
of motion - because the parameters appear only in the combination $q$
and clearly modifications of the bulk action are needed to see
``subleading'' effects.

We begin with the order parameter by keeping the operator undergoing
condensation (i.e., $\D$) fixed and varying the coupling constant
$q$. 

The accompanying graph clearly shows the difference in the two
families of condensates, there being an order of magnitude difference
between $\D=1$ and $\D=2$ (as observed earlier in \cite{KKY}). We have
also overlaid best-fit curves $\langle{\mathcal O}_0\rangle\sqrt{1-x^\nu}$. 
\begin{figure}[H]
  \centering
\includegraphics[height=6cm]{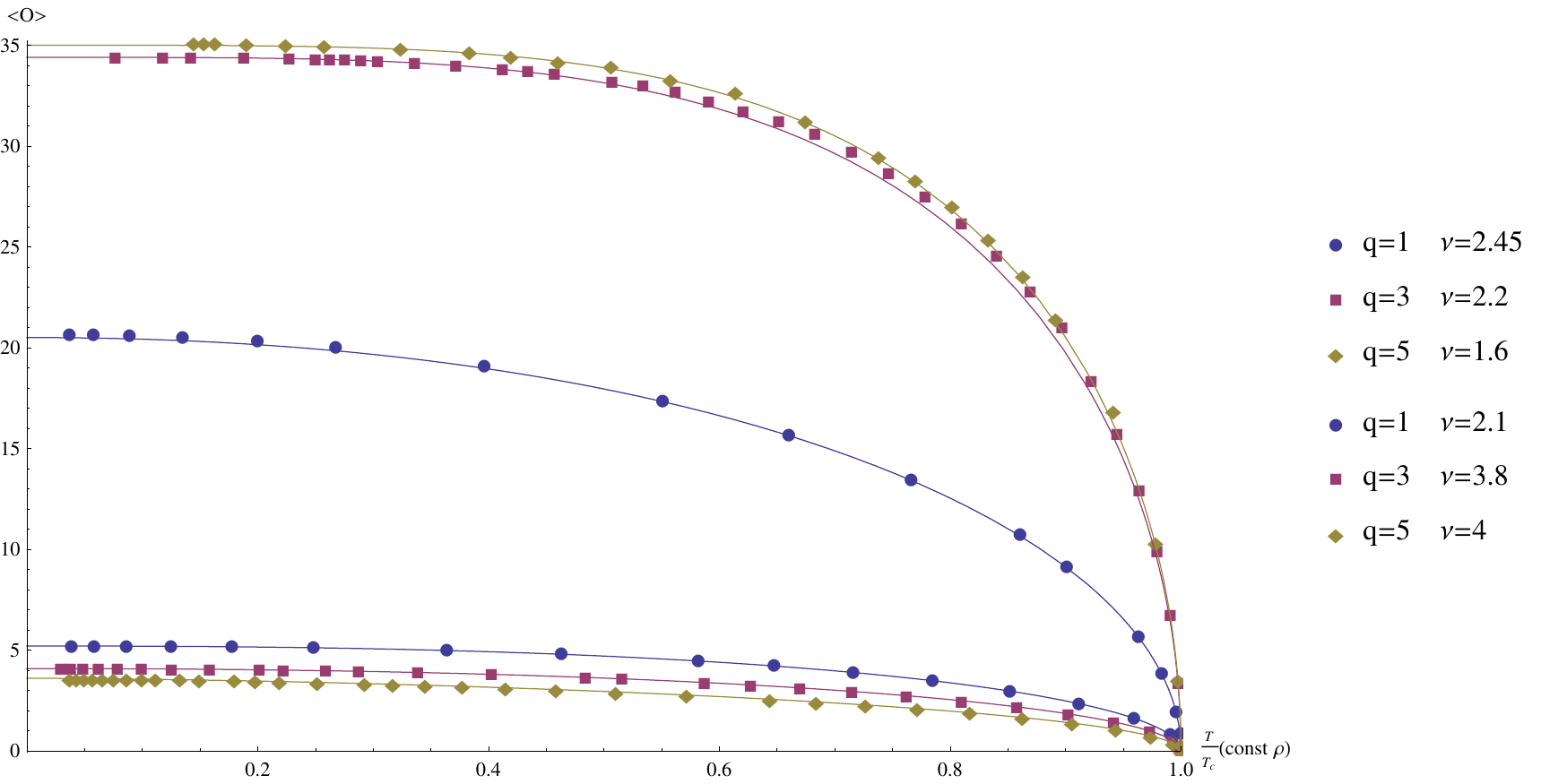}
\caption{\label{vevTq} $\frac{<{\mathcal O}>}{{\mathcal O_0}}$ vs
  $\frac{T}{T_c}$ at constant density $\rho$}
\end{figure}

The larger values of $q$ are closer to the probe limit when the scalar
field backreaction can be safely ignored. The probe limit is however
attained differently for $\D>\frac{3}{2}$ compared to $\D<\frac{3}{2}$
(this difference in the behaviour as a function of $q$ was already
noted by \cite{HHH2}). Both the exponent and the amount of
condensation $\mathcal{O}_0$ decrease with $q$ for $\D<\frac{d}{2}$
but increase for larger $\D$.

Putting together the story for various other values of $\D$, we get
the following Table: \ref{Tableq} where the trend in both the exponent
and the magnitude of the VEV are clearly seen as a function of $q$.
\begin{center}
  \begin{tabular}{|l|c|c|c||c|c|c|}
  \hline
$\D\downarrow \, q \rightarrow$ & 1 & 3 & 5 & 1&3&5\\\hline
  1            &2.45 & 2.2  & 1.6& 5.2  &4.08  &3.6   \\ \hline
 $\frac{5}{4}$ &2.6  & 2.8  & 2.3& 4.42 &3.67  &3.27   \\ \hline
 $\frac{7}{4}$ &2.3  & 3.7  & 3.5& 8.9  &10.5  &10   \\ \hline
  2            &2.1  & 3.8  & 4  & 20.5 & 34.4 & 35     \\ \hline
  \end{tabular}
\label{Tableq}
\end{center}

The first part of the table show the behaviour of the exponent while
the second part shows the behaviour of the zero temperature value
$\mathcal{O}_0$ in
$\langle\mathcal{O}\rangle=\mathcal{O}_0\sqrt{1-\left(\frac{T}{T_c}\right)^\n
}$. For any q, it is noteworthy that $\mathcal{O}_0$ for $\D=2$ is an
order of magnitude larger than that for $\D=1$.

The energy density continues to behave as $T^3$ (at fixed $\r$) for
all values of $q$ except that for larger values of $\D$, the energy
density goes to nearly zero at $T\to 0$. Correspondingly, the
exponents in the entropy vs. density graph (at constant energy) all
remain the same at $5/8$ for all $q$ and all $\D$. 


The entropy density which was a power law of temperature (near
$T=T_c$) continues to be a power law. However, for larger values of
$q$ - the exponents which varied significantly for $q=1$ (see Table:
\ref{Table}) become nearly identical at $\a=2.25$ (is this the exponent
for the black hole?). Of course in the probe limit, the condensation
process does not affect the black hole and hence the specific heat
becomes continuous as well.


In the case of the entropy as a function of the density (at fixed
temperature or thermal density) - we have seen that the entropy
decreases initially for the smaller $\D$ (Fig: \ref{sqT}). For larger
values of $q\geq 2$, the minima disappears, and all curves maybe
simply described as a simple power law $s\sim \r^{\frac{3}{4}}$ (for
all $\D$).
 
\section{Free energy and Entropy}

All along, we have identified the entropy of the dual system with the
area of the horizon of the black hole $s=\frac{1}{4\pi z_H ^2}$ . On
the other hand, the entropy of the dual system may also be evaluated
by considering the Grand Potential $\O$ and taking its derivative with
respect to $T$--$s=(\frac{\del \O}{\del T})_\mu$ . It will be
interesting to ask if these two values agree. In the holographic
language, taking derivatives of the grand potential requires us to
compare two classical solutions which is similar to considering a
(quasi-static) thermodynamic process. Put this way, it is quite
non-obvious that these two evaluations will give us the same result,
especially because of the presence of the condensate.

\begin{figure}
  \centering
  \includegraphics[width=0.6\textwidth]{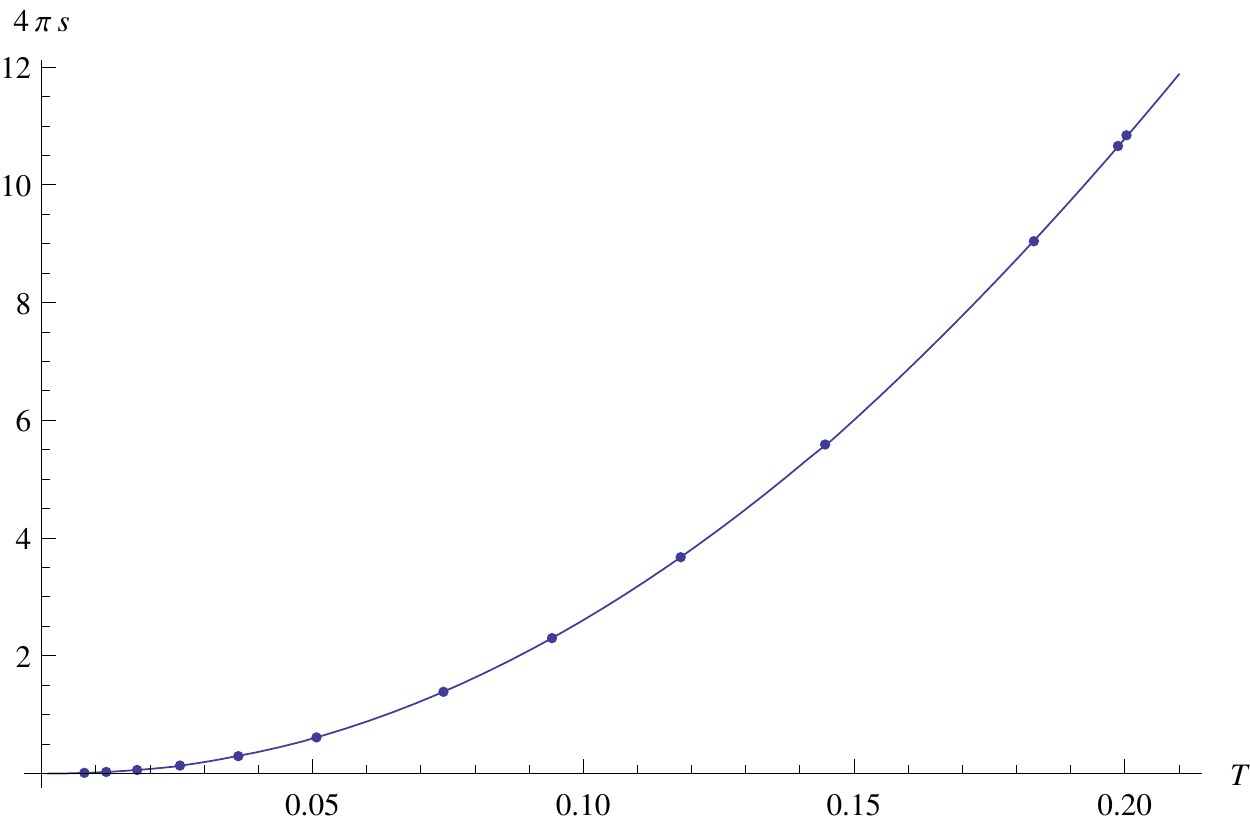}
  \caption{$4\pi s$ vs. T at constant $\mu$ for $q=1,\D=1$  }
  \label{checksT}
\end{figure}

To check this, we first use an interpolation function to fit the
values $\O(T)$ for fixed chemical potential. The numerical derivative
$s=(-\frac{\del \O}{\del T})_\mu$ is plotted as a solid curve in the
figure \ref{checksT} against the entropy data obtained from individual
solutions as $s=\frac{1}{4\pi z_H ^2}$ shown as dots- showing
excellent agreement.

\begin{figure}[H]
  \centering
  \includegraphics[width=0.6\textwidth]{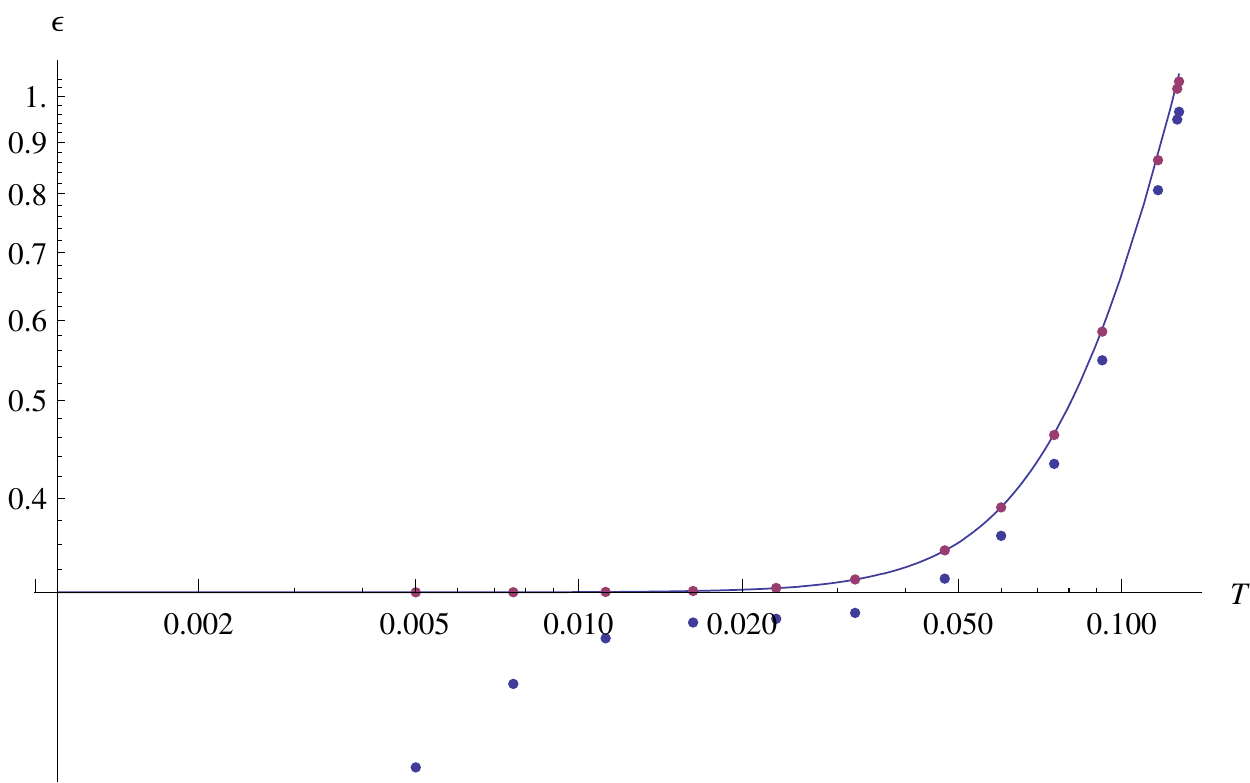}
  \caption{$e$ vs. T at constant $\mu$ for $q=1,\D=1$  }
  \label{checkeT}
\end{figure}

Another such thermodynamic relation is $\frac{\del
  (\beta\Omega)}{\del\beta}=U=T_{00}$ - which we show in figure
\ref{checkeT}. In this graph, we show the energy data points obtained
from using the FG-expansion $\e=-\frac{d}{4\pi G_N} \frac{2}{3}
g^{(d)} _{00}$ (blue dots) as described earlier - along with the
energy values determined by using the Euler relation (magenta colored
dots). These are shown against the curve obtained by using an
interpolating function and differentiating.

These equalities are expected if we identify $\log Z=-\beta\Omega$
with the grand partition function. Many other such thermodynamic
consistency conditions and relations have been studied/used in works
on holographic fluid dynamics and also on holographic sum rules. For
the charged black hole alone - these relations can be verified
analytically as detailed in the appendix.

\section{Discussion}

In this paper - we have assembled together a fairly comprehensive
treatment of the thermostatic properties of a particular family of
holographic superfluid systems. We have shown that upon inclusion of
the condensate solutions, a complete phase diagram can be
obtained. Here, we focused on the phenomenological differences that
occur due to varying scaling dimensions. We have presented a careful
exploration of thermodynamic consistency of the holographic recipe -
by showing that familiar thermodynamic relations are indeed obeyed by
these holographic systems.

A utility of the holographic description is that some of the features
of the field theory thermodynamics maybe understood as a competition
between gravitational attraction and electric repulsion between the
black hole and the charge cloud of the scalar field. Another approach
via the Raychaudhuri equations \cite{Tian} on the gravitational side
could lead to further insights on the behaviour of the field theory
system - especially from the viewpoint of the quantum to thermal
transition that happens when the condensate occupancy is varied
keeping the total number of particles fixed. This line of exploration
will be able to shed some further light on the condensation of neutral
scalars as well.

The dependence of various thermostatic quantities compare favorably
with those obtained from experimental systems probing the physics of
gases of cold atoms. These comparisons must be approached with
caution - since there is scant motivation. However, the holographic
graphs do seem to capture interaction effects going well beyond
perturbation theory and even mean field theory. Thus, a careful and
more comprehensive comparison in four boundary dimensions, this time
including transport properties and relaxation time scales might be an
exciting place to confront holography with experiments.

In fact, our study has already suggested that isentropic processes
maybe able to better capture the effect of the interactions and
highlight qualitative differences as we vary the interaction
strength. We have also shown that the critical exponent of the order
parameter depends on the bulk gauge charge apart from the nature of
the order parameter (quantified by the scaling dimension). This latter
feature is relevant for further exploration along the lines of
holographic QCD equations of state since the gauge field in the bulk
gets reinterpreted as giving rise to a flavor symmetry in the
boundary. We have also pointed out that for the purpose of
phenomenology of these systems, additional (bulk) interactions that
break the degeneracy among the parameters $\l,e,G_N$ will be required.
The manner in which $q$ appears in the thermodynamic quantities in the
main text suggest specific additional bulk couplings which might be
expected to play a significant role (in fact, one should re-examine
the literature surveyed in the introduction in this light since
significant work has already been done along these directions).
We hope to explore a few of these questions in the near future. 


\section{Charged black hole}

In this section, we summarize the thermodynamics of charged black
holes in $AdS_{d+1}$ space.

\begin{eqnarray}
\Psi(z)=0 \hspace{1cm} \chi(z)=4 \log z\\
\phi(z)=\mu+Q z^{d-2} \\
g(z)=z^2+\frac{(d-2) Q^2 z^{2d}}{2 q^2(d-1)}-c z^{2+d}
\end{eqnarray}
The parameters $q, Q$ and $c$ determine the temperature of the
AdS-Schwarzschild black hole. For positive $c$, is this solution
unstable in the usual thermodynamic sense. Note that $c$ also has the
interpretation of energy in the dual theory while the condition that
the scalar potential vanish at the horizon determines the chemical
potential to be $\mu=-\frac{Qz_0 ^{d-2}}{q}$ where $Q$ is related to the number density $\rho$ as $\rho=\frac{(d-2)Q z_H ^{d-3}}{16\pi q G_N z_H ^{d-1}}$

By rescaling $r=\lambda z$ and defining $\alpha=\frac{(d-2) Q^2}{2 q^2(d-1)}$,  $$g=\lambda^2 r^2(1-c
\lambda^d(r^{d}-\frac{\alpha \lambda^{d-2}}{c}r^{2d-2}))$$ If we
choose $\frac{\alpha \lambda^{d-2}}{c}=1$, this fixes $\lambda$ and
the function can be rewritten as
$$g=r^2 c\lambda^{d+2}(\beta-r^{d}+r^{2d-2})$$ 
with $\beta=\frac{1}{c\lambda^d}$ and hence the root $r_0(\beta)$ is a
function of the combination $\beta$ alone.
The temperature is determined from the conical singularity argument to be
\begin{equation}
  T(c,\frac{Q}{q})=\frac{g'[r_0]\exp[-\frac{\chi}{2}]}{4\pi}
  =\frac{c\lambda^{d-1}}{4\pi r_0}(2\beta-(d+2)r_0 ^{d}+2dr_0 ^{2d-2}))
\end{equation}

From the expression for the location of the horizon, we can see that
there is a condition $0<\beta<\frac{27}{256} (d=3)$ which in turn
means that there is an upper bound on $Q$ given $c,q$. That is to say,
for a given energy $c$, the charge density that the condensate free
``black hole'' can sustain has a limit. This statement can also be
translated into the canonical ensemble and maybe said to be the dual
of the statement that there is a maximum number density that can be
supported by the excited states of a bosonic system at finite
temperature (or energy). Increasing the density further could lead to
BEC - thus finite occupancy of the ground state. In the AdS side, this
means that we have a charged scalar field with a nonvanishing profile.


From the analytic solution, we may extract the following boundary
data.
\begin{eqnarray}
  \e=\frac{(d-1)}{16\pi G_N z_H ^d}(1+\a z_H ^{2d-2})
  \quad P=\frac{\e}{d-1} \quad \r=\frac{1}{16\pi G_N} \frac{(d-2)Q}{q}  \\
  s=\frac{1}{4 G_Nz_H ^{d-1}}\quad T=\frac{1}{4\pi z_H}(d-(d-2)\a z_H ^{2d-2})
  \quad \m=-\frac{Q}{q z_H ^{d-2}}\\
\end{eqnarray}

From this data, we can see that the Euler relation $\frac{\m \r + T
  s}{\e+P}=1$ is trivially satisfied.


\begin{thebibliography}{999}
\bibitem{Gubser} S.~S.~Gubser,
  Phys.\ Rev.\ D {\bf 78}, 065034 (2008)
  [arXiv:0801.2977 ].
  \bibitem{HHH} 
  S.~A.~Hartnoll, C.~P.~Herzog and G.~T.~Horowitz,
  Phys.\ Rev.\ Lett.\  {\bf 101}, 031601 (2008)
  [arXiv:0803.3295 ].
\bibitem{HHH2}
  S.~A.~Hartnoll, C.~P.~Herzog and G.~T.~Horowitz,
  JHEP {\bf 0812}, 015 (2008)
  [arXiv:0810.1563 ].
  
\bibitem{Hartnoll:2009sz} 
  S.~A.~Hartnoll,
  Class.\ Quant.\ Grav.\  {\bf 26}, 224002 (2009)
  [arXiv:0903.3246 ].

\bibitem{McGreevy:2009xe} 
  J.~McGreevy,
  Adv.\ High Energy Phys.\  {\bf 2010}, 723105 (2010)
  [arXiv:0909.0518 ].

\bibitem{Hartnoll:2011fn} 
  S.~A.~Hartnoll,
  arXiv:1106.4324 .

\bibitem{Sachdev:2011wg} 
  S.~Sachdev,
  Ann.\ Rev.\ Condensed Matter Phys.\  {\bf 3}, 9 (2012)
  [arXiv:1108.1197 [cond-mat.str-el]].

  
\bibitem{Adams:2012th} 
  A.~Adams, L.~D.~Carr, T.~Schäfer, P.~Steinberg and J.~E.~Thomas,
  New J.\ Phys.\  {\bf 14}, 115009 (2012)
  [arXiv:1205.5180 ].

\bibitem{Musso:2014efa} 
  D.~Musso,
  PoS Modave {\bf 2013}, 004 (2013)
  [arXiv:1401.1504 ].

\bibitem{Cai:2015cya} 
  R.~G.~Cai, L.~Li, L.~F.~Li and R.~Q.~Yang,
  Sci.\ China Phys.\ Mech.\ Astron.\  {\bf 58}, no. 6, 060401 (2015)
  [arXiv:1502.00437 ].


\bibitem{McGreevy:2016myw} 
  J.~McGreevy,
  arXiv:1606.08953 .
\bibitem{what}
  Stefano Giorgini, Lev P. Pitaevskii, Sandro Stringari,
%
  Rev.\ Mod.\ Phys. {\bf 80}, 1215 
  [arXiv:0706.3360]

\bibitem{Sunethra}
  S.~Ramanan and M.~Urban,
  Phys.\ Rev.\ C {\bf 88}, no. 5, 054315 (2013)
  [arXiv:1308.0939 [nucl-th]].
\bibitem{Stringari}
M.~Antezza, F.~Dalfovo, L.~P.~Pitaevskii, S.~Stringari
Phys.\ Rev.\ A {\bf 76}, 043610 (2007)
	[arXiv:0706.0601].

\bibitem{Luo:PRL}
Luo, L.; Clancy, B.; Joseph, J.; Kinast, J.; Thomas, J. E.
Physical Review Letters, vol. 98, Issue 8, id. 080402
[arXiv:cond-mat/0611566]

\bibitem{Skenderis} 
S.~de Haro, S.~N.~Solodukhin and K.~Skenderis,
  Commun.\ Math.\ Phys.\  {\bf 217} (2001) 595
  [hep-th/0002230]

\bibitem{Chamblin}

  A.~Chamblin, R.~Emparan, C.~V.~Johnson and R.~C.~Myers,
  Phys.\ Rev.\ D {\bf 60}, 104026 (1999)
  [hep-th/9904197].
  
\bibitem{Winters}
  M.~Kruczenski, D.~Mateos, R.~C.~Myers and D.~J.~Winters,
  JHEP {\bf 0307}, 049 (2003)
  [hep-th/0304032].
  
\bibitem{KW}
I.~R.~Klebanov and E.~Witten,
  Nucl.\ Phys.\ B {\bf 556}, 89 (1999)
  [hep-th/9905104].













\bibitem{Horowitz:2008bn} 
  G.~T.~Horowitz and M.~M.~Roberts,
  Phys.\ Rev.\ D {\bf 78}, 126008 (2008)
  [arXiv:0810.1077 ].

\bibitem{Gubser:2008pf} 
  S.~S.~Gubser and A.~Nellore,
  JHEP {\bf 0904}, 008 (2009)
  [arXiv:0810.4554 ].

\bibitem{Gubser:2009cg} 
  S.~S.~Gubser and A.~Nellore,
  Phys.\ Rev.\ D {\bf 80}, 105007 (2009)
  [arXiv:0908.1972 ].

\bibitem{Horowitz:2009ij} 
  G.~T.~Horowitz and M.~M.~Roberts,
  JHEP {\bf 0911}, 015 (2009)
  [arXiv:0908.3677 ].

\bibitem{SJS}
  Y.~Kim, Y.~Ko and S.~J.~Sin,
  Phys.\ Rev.\ D {\bf 80}, 126017 (2009)
  [arXiv:0904.4567 ].

  
\bibitem{Umeh:2009ea} 
  O.~C.~Umeh,
  JHEP {\bf 0908}, 062 (2009)
  [arXiv:0907.3136 ].

\bibitem{Franco:2009if} 
  S.~Franco, A.~M.~Garcia-Garcia and D.~Rodriguez-Gomez,
  Phys.\ Rev.\ D {\bf 81}, 041901 (2010)
  [arXiv:0911.1354 ].

  
\bibitem{KKY1} 
  V.~Keranen, E.~Keski-Vakkuri, S.~Nowling and K.~P.~Yogendran,
  Phys.\ Rev.\ D {\bf 81}, 126011 (2010)
  [arXiv:0911.1866 ].

\bibitem{KKY2} 
  V.~Keranen, E.~Keski-Vakkuri, S.~Nowling and K.~P.~Yogendran,
  Phys.\ Rev.\ D {\bf 81}, 126012 (2010)
  [arXiv:0912.4280 ].
  
\bibitem{Aprile:2009ai} 
  F.~Aprile and J.~G.~Russo,
  Phys.\ Rev.\ D {\bf 81}, 026009 (2010)
  [arXiv:0912.0480 ].

\bibitem{Pan:2009xa} 
  Q.~Pan, B.~Wang, E.~Papantonopoulos, J.~Oliveira and A.~B.~Pavan,
  Phys.\ Rev.\ D {\bf 81}, 106007 (2010)
  [arXiv:0912.2475 ].

\bibitem{Brihaye:2010mr} 
  Y.~Brihaye and B.~Hartmann,
  Phys.\ Rev.\ D {\bf 81}, 126008 (2010)
  [arXiv:1003.5130 ].

\bibitem{Liu:2010ka} 
  Y.~Liu and Y.~W.~Sun,
  JHEP {\bf 1007}, 099 (2010)
  [arXiv:1006.2726 ].
  

\bibitem{Charmousis:2010zz} 
  C.~Charmousis, B.~Gouteraux, B.~S.~Kim, E.~Kiritsis and R.~Meyer,
  JHEP {\bf 1011}, 151 (2010)
  [arXiv:1005.4690 ].

\bibitem{Arean:2010zw} 
  D.~Arean, P.~Basu and C.~Krishnan,
  JHEP {\bf 1010}, 006 (2010)
  [arXiv:1006.5165 ].

\bibitem{Way}
  G.~T.~Horowitz and B.~Way,
  JHEP {\bf 1011}, 011 (2010)
  [arXiv:1007.3714 ].

\bibitem{KKY}
  V.~Keranen, E.~Keski-Vakkuri, S.~Nowling and K.~P.~Yogendran,
  New J.\ Phys.\  {\bf 13}, 065003 (2011)
  [arXiv:1012.0190 ].

\bibitem{Edalati:2010ge} 
  M.~Edalati, R.~G.~Leigh, K.~W.~Lo and P.~W.~Phillips,
  Phys.\ Rev.\ D {\bf 83}, 046012 (2011)
  [arXiv:1012.3751 ].

\bibitem{Pan:2011ns} 
  Q.~Pan and B.~Wang,
  arXiv:1101.0222 .

  
\bibitem{Dias:2011tj} 
  O.~J.~C.~Dias, P.~Figueras, S.~Minwalla, P.~Mitra, R.~Monteiro and J.~E.~Santos,
  JHEP {\bf 1208}, 117 (2012)
  [arXiv:1112.4447 ].


\bibitem{Liu:2014mva} 
  Y.~Liu, K.~Schalm, Y.~W.~Sun and J.~Zaanen,
  JHEP {\bf 1405}, 122 (2014)
  [arXiv:1404.0571 ].
  
  


\bibitem{Arean:2015wea} 
  D.~Arean and J.~Tarrio,
  JHEP {\bf 1504}, 083 (2015)
  [arXiv:1501.02804 ].

 

\bibitem{Plantz:2015pem} 
  N.~W.~M.~Plantz, H.~T.~C.~Stoof and S.~Vandoren,
  arXiv:1511.05112 .
  
  

\bibitem{Chen:2016cym} 
  J.~W.~Chen, S.~H.~Dai, D.~Maity and Y.~L.~Zhang,
  Phys.\ Rev.\ D {\bf 94}, no. 8, 086004 (2016)
  [arXiv:1603.08259 ].



\bibitem{deWolfe}
  O.~DeWolfe, O.~Henriksson and C.~Wu,
  arXiv:1611.07023 .


\bibitem{Karch}
  A.~Karch and B.~Robinson,
  JHEP {\bf 1512}, 073 (2015)
  [arXiv:1510.02472 ].


\bibitem{Yin:2015} 
Lei Yin, Defu Hou, Hai-cang Ren
Phys. Rev. D {\bf 91}, no. 2, 026003 (2015)
[arXiv:1311.3847 ].




  
\bibitem{Sonner:2010yx} 
  J.~Sonner and B.~Withers,
  Phys.\ Rev.\ D {\bf 82}, 026001 (2010)
  [arXiv:1004.2707 ].


\bibitem{Kiritsis:2012ma} 
  E.~Kiritsis and V.~Niarchos,
  JHEP {\bf 1208}, 164 (2012)
  [arXiv:1205.6205 ].


\bibitem{Amado:2013aea} I.~Amado, D.~Arean, A.~Jimenez-Alba,
  K.~Landsteiner, L.~Melgar and I.~Salazar Landea,
  JHEP {\bf 1402}, 063 (2014)
  [arXiv:1307.8100 ].
  
\bibitem{Sonner:2014tca} 
  J.~Sonner, A.~del Campo and W.~H.~Zurek,
  Nature Commun.\  {\bf 6}, 7406 (2015)
  [arXiv:1406.2329 ]

  
  
\bibitem{Iqbal:2011bf} 
  N.~Iqbal and H.~Liu,
  Class.\ Quant.\ Grav.\  {\bf 29}, 194004 (2012)
  [arXiv:1112.3671 ].

\bibitem{Hartnoll:2012pp} 
  S.~A.~Hartnoll and R.~Pourhasan,
  JHEP {\bf 1207}, 114 (2012)
  [arXiv:1205.1536 ]

\bibitem{Ahn:2015shg} 
  B.~Ahn, S.~Hyun, K.~K.~Kim, S.~A.~Park and S.~H.~Yi,
  Phys.\ Rev.\ D {\bf 94}, no. 2, 024043 (2016)
  [arXiv:1512.09319 ].


\bibitem{Tian}
    Y.~Tian, X.~N.~Wu and H.~B.~Zhang,
  ``Holographic Entropy Production,''
  JHEP {\bf 1410}, 170 (2014)
  [arXiv:1407.8273 ].


\bibitem{Ku}
  Mark J. H. Ku, Ariel T. Sommer, Lawrence W. Cheuk, Martin W. Zwierlein
  ``Revealing the Superfluid Lambda Transition in the Universal Thermodynamics of a Unitary Fermi Gas,''
  Science 335, 563-567 (2012)
  [arXiv:1110.3309]









  
\end{thebibliography}
\end{document}